\documentclass[twoside,draft,reqno]{birkart}
\usepackage{amssymb}
\renewcommand{\theequation}{\thesection.\arabic{equation}}
\renewcommand{\ll}{\label} \newcommand{\begeq}{\begin{equation}}
\newcommand{\bea}{\begin{eqnarray}}
\newcommand{\eea}{\end{eqnarray}} \newcommand{\nn}{\nonumber}
 \newcommand{\ci}{\cite}
 
\newcommand{\ca}{$C^*$-algebra} 
 \newcommand{\rep}{representation}

\newcommand{\Hs}{Hilbert space}

   \newcommand{\vna}{von
Neumann algebra}
\newcommand{\op}{^{\mbox{\tiny op}}}
\newcommand{\id}{\mbox{\rm id}}

 \newcommand{\ovl}{\overline}
 \newcommand{\til}{\tilde}
\newcommand{\raw}{\rightarrow} 
\newcommand{\rac}{\rightarrowtail}
\newcommand{\lac}{\leftarrowtail}
\newcommand{\law}{\leftarrow} \newcommand{\Raw}{\Rightarrow}
\newcommand{\hraw}{\hookrightarrow} \newcommand{\Law}{\Leftarrow}
\newcommand{\lraw}{\leftrightarrow}

\newcommand{\rlh}{\rightleftharpoons} \newcommand{\n}{\|}
\newcommand{\ot}{\otimes} 
\newcommand{\la}{\langle} \newcommand{\ra}{\rangle}

\newcommand{\x}{\times} 
 
\newcommand{\Tr}{\mbox{\rm Tr}\,} 
\newcommand{\Co}{{\rm Co}}

 \newcommand{\BH}{\mathfrak{B}({\mathcal H})} 
\newcommand{\cin}{C^{\infty}} \newcommand{\cci}{C^{\infty}_c}

 \newcommand{\Hlg}{{\mathcal H}_{\chi}}

\newcommand{\inv}{^{-1}} 
\newcommand{\Exp}{{\rm Exp}}

\newcommand{\bb}{\rangle_{\mathfrak{B}}}

\newcommand{\al}{\alpha} 
\newcommand{\gm}{\gamma} \newcommand{\Gm}{\Gamma}

\newcommand{\io}{\iota} 
\newcommand{\lm}{\lambda} 
\newcommand{\rh}{\rho} \newcommand{\sg}{\sigma}
\newcommand{\Sg}{\Sigma} \newcommand{\ta}{\tau} 
\newcommand{\Ph}{\Phi} \newcommand{\phv}{\varphi}
\newcommand{\ch}{\chi} \newcommand{\ps}{\psi} \newcommand{\Ps}{\Psi}
\newcommand{\om}{\omega} \newcommand{\Om}{\Omega}
\newcommand{\Up}{\Upsilon}
\newcommand{\A}{\mathfrak{A}} \newcommand{\B}{\mathfrak{B}}
\newcommand{\GC}{\mathfrak{C}}

\newcommand{\GM}{\mathfrak{M}} \newcommand{\GN}{\mathfrak{N}}

\newcommand{\h}{\mathfrak{h}} 
 
\newcommand{\GP}{\mathfrak{P}} 

 \newcommand{\CF}{{\mathcal F}}
\newcommand{\CD}{{\mathcal D}} \newcommand{\CE}{{\mathcal E}}
 \renewcommand{\H}{{\mathcal H}}
 
\newcommand{\CK}{{\mathcal K}}   \newcommand{\CL}{{\mathcal L}}   
 \newcommand{\CM}{{\mathcal M}}
\newcommand{\CN}{{\mathcal N}} \newcommand{\CS}{{\mathcal S}}
\newcommand{\CO}{{\mathcal O}}

\renewcommand{\L}{{\mathcal L}}
\newcommand{\C}{{\mathbb C}} 
 \newcommand{\I}{{\mathbb I}}
 \newcommand{\R}{{\mathbb R}}
 
\newcommand{\SSS}{\mathsf{S}} %
  \makeatletter
\newskip\tempskip \def\endproof{{\parfillskip24\p@ plus\@ne
fil\@@par}\tempskip\prevdepth
\ifdim\lastskip=\z@\tempskip\z@\else\vskip-\lastskip
\ifdim\tempskip>4\p@ \tempskip.5\tempskip \else \tempskip\z@\fi\fi
\nobreak\vskip-\baselineskip\vskip-\tempskip\noindent\hbox
to\hsize{\hfill
$\blacksquare$}\par\vskip\tempskip\vskip\abovedisplayskip\@doendpe}
\makeatother \makeatletter
\newskip\tempskip \def\endiproof{{\parfillskip24\p@ plus\@ne
fil\@@par}\tempskip\prevdepth
\ifdim\lastskip=\z@\tempskip\z@\else\vskip-\lastskip
\ifdim\tempskip>4\p@ \tempskip.5\tempskip \else \tempskip\z@\fi\fi
\nobreak\vskip-\baselineskip\vskip-\tempskip\noindent\hbox
to\hsize{\hfill
$\Box$}\par\vskip\tempskip\vskip\abovedisplayskip\@doendpe}
\makeatother 

\newcommand{\otB}{\hat{\otimes}_{\mathfrak{B}}}
\newcommand{\otA}{\hat{\otimes}_{\mathfrak{A}}}
\newcommand{\otq}{\hat{\otimes}}
\newcommand{\otc}{\circledcirc}
\newcommand{\otg}{\circledast}

\newcommand{\Me}{Morita equivalent}

\newcommand{\Rep}{\mbox{\rm Rep}}
\newcommand{\End}{\mbox{\rm End}}
\newcommand{\Hom}{\mbox{\rm Hom}}

\newcommand{\Ca}{\mbox{\rm \textsf{C}$\mbox{}^*$}}
\newcommand{\Wa}{\mbox{\rm \textsf{W}$\mbox{}^*$}}
\newcommand{\Gr}{\mbox{\rm \textsf{G}}}
\newcommand{\Grb}{\mbox{\rm \textsf{G'}}}
\newcommand{\LG}{\mbox{\rm \textsf{LG}}}
\newcommand{\MG}{\mbox{\rm \textsf{MG}}}

\newcommand{\SyG}{\mbox{\rm \textsf{SG}}}
\newcommand{\Po}{\mbox{\rm \textsf{Poisson}}}
\newcommand{\Alg}{\mbox{\rm \textsf{Alg}}}
\newcommand{\LGb}{\mbox{\rm \textsf{LG'}}}
\newtheorem{theorem}{Theorem}[section]
\newtheorem{lemma}[theorem]{Lemma}
\newtheorem{corollary}[theorem]{Corollary}
\newtheorem{proposition}[theorem]{Proposition}

\newtheorem{remark}[theorem]{Remark}
\newtheorem{example}[theorem]{Example}

\newtheorem{definition}[theorem]{Definition}

\newenvironment{Proof}{\removelastskip\par\medskip
\noindent{\em Proof.} \rm}{\penalty-20\null\hfill$\square$\par\medbreak}

\begin{document}
\setlength{\unitlength}{1cm}

\title[Quantized symplectic reduction]
{Quantized reduction as a tensor product}
\author[N.P. Landsman]{N.P. Landsman}

\address{
Korteweg--de Vries Institute for Mathematics,\br
University of Amsterdam,\br
Plantage Muidergracht 24,\br
NL-1018 TV AMSTERDAM, THE NETHERLANDS}

\email{npl@science.uva.nl}

\begin{abstract}
Symplectic reduction is reinterpreted as the composition of arrows in
the category of integrable Poisson manifolds, whose arrows are
isomorphism classes of dual pairs, with symplectic groupoids as units.
Morita equivalence of Poisson manifolds amounts to isomorphism of objects
in this category.

This description paves the way for the quantization of the classical
reduction procedure, which is based on the formal analogy between dual
pairs of Poisson manifolds and Hilbert bimodules over \ca s, as well
as with correspondences between
\vna s.  Further analogies are drawn with categories of  groupoids
(of algebraic, measured, Lie, and symplectic type).  In all cases, the
arrows are isomorphism classes of  appropriate bimodules, and their
composition may be seen as a tensor product.  Hence in suitable
categories reduction is simply composition of arrows, and Morita
equivalence is isomorphism of objects.
\end{abstract}
\maketitle
\tableofcontents
\section{Introduction}
  In a formalism where classical and quantum mechanics are described
  by Poisson manifolds and \ca s,  respectively, the theory of
  constrained quantization can be developed on the basis of the
  analogy between dual pairs of Poisson
  manifolds on the one hand, and Hilbert bimodules over \ca s (or
  correspondences of
\vna s) on the other. On the classical side, a
 dual pair $Q\stackrel{q}{\law} S
\stackrel{p}{\raw}P$ of two Poisson manifolds $P$, $Q$ consists of a
symplectic space $S$ with complete Poisson maps $p:S\raw P^-$ and
$q:S\raw Q$, such that $\{p^*f,q^*g\}=0$ for all $f\in\cin(P)$ and
$g\in\cin(Q)$ \cite{K1,W1}.  Under suitable regularity conditions, one may define a
``tensor product'' $\otc_P$ between a $Q$-$P$ dual pair and a $P$-$R$
dual pair, yielding a $Q$-$R$ dual pair.  This tensor product may
equivalently be defined either by symplectic reduction or through a
construction involving symplectic groupoids.  In other words, symplectic reduction,
including the special case of Marsden--Weinstein--Meyer  reduction, may be
formulated as the tensor product of suitable dual pairs.
 
On the quantum side, an $\A$-$\B$ Hilbert bimodule $\A\rac\CE\rlh\B$,
where $\A$ and $\B$ are \ca s, consists of a complex Banach space
$\CE$ that is an algebraic $\A$-$\B$ bimodule, and is equipped with a
$\B$-valued inner product that is compatible with the $\A$ and $\B$
actions. There exists an ``interior'' tensor product $\hat{\ot}_{\B}$
of an $\A$-$\B$ Hilbert bimodule with a $\B$-$\GC$ Hilbert bimodule,
first defined by Rieffel \cite{Rie1,Rie2}, producing an $\A$-$\GC$
Hilbert bimodule. This ``quantum'' tensor product quantizes
the ``classical'' one mentioned above \cite{NPL1}. 
An appropriate choice of $\B$ (namely $\B=C^*(H)$, a group
\ca) then provides a quantum version of Marsden-Weinstein reduction,
which for compact groups turns out to be equivalent to Dirac's
well-known method for quantizing first-class constraints \cite{NPL3}.

When $\A$ and $\B$ are \vna s, the notion of a Hilbert bimodule
may be replaced by that of a correspondence \cite{Con}, which is
easier to define and work with. A correspondence between
two \vna s $\GM,\GN$ is simply a \Hs\ on which $\GM$ acts from the
left  and $\GN$ acts from the right, such that the actions commute
(and each action is normal, i.e., $\sg$-weakly continuous).
A \vna ic analogue of Rieffel's interior tensor product,
sometimes called the relative tensor product \cite{Sau} or Connes
fusion \cite{Was}, has been defined by Connes \cite{Con}. 
This may be seen as a quantization of the classical tensor product as
well.

The analogy between dual pairs on the one hand, and Hilbert
bimodules or correspondences on the other, is well illustrated by the
theory of Morita equivalence. This theory originally applied to algebras;
cf.\ \cite{faith}.  A key theorem is that two
algebras have equivalent categories of left (or right) modules iff they
are related by a bimodule with certain properties, called an
equivalence bimodule. Such algebras, then, are called Morita equivalent.
 The proof of this theorem crucially relies on
the usual (algebraic) bimodule tensor product.

A version of Morita's theory appropriate for \ca s was developed by
Rieffel \cite{Rie1,Rie2}. Basically, the \ca ic theory looks like the
one for algebras, with modules, bimodules, and the algebraic bimodule
tensor product replaced by Hilbert spaces, Hilbert bimodules, and
Rieffel's interior tensor product, respectively.  For \vna s one here
has Hilbert spaces, correspondences, and Connes's relative tensor
product. Similarly, the theory of Morita equivalence for Poisson
manifolds, initiated by Xu \cite{X2}, involves symplectic manifolds,
dual pairs, and the ``classical'' tensor product mentioned above.

 Groupoids provide a further area of application of tensor products
 and Morita equivalence. Further motivation for their study comes from
 the central role they play in the description of singular spaces,
 such as the leaf space of a foliation \cite{Con,MoS}, or a manifold
 with singularities \cite{LauNis}.  Lie groupoids are important in
 quantization theory \cite{NPL3,CW}, and symplectic groupoids have
 revolutionized Poisson geometry \cite{K0,KM,W3,CDW,MiWe,Zak}. For our
 purposes, symplectic groupoids are firstly needed to define
 regularity conditions on dual pairs that guarantee the existence of
 the classical tensor product, and secondly play the role of unit arrows in
the category of Poisson manifolds.

 Our presentation will be based on the idea that the bimodules of
various sorts are best seen as arrows in a category, and that each of
the tensor products under consideration simply defines the composition
of matched arrows. Since each of our tensor products is merely
associative up to isomorphism, and admits unit objects also merely up
to isomorphism, this idea has to be implemented in one of the
following two ways.  Either one works with bicategories rather than
categories, as done in \cite{NPLDR}, or one defines arrows as
isomorphism classes of bimodules. These options are not equivalent;
the bicategorical approach is richer in structure, but the categorical
one is easier to handle. Therefore, in the present
paper we take the second route. 

Algebras form the objects of a category \Alg, whose arrows are isomorphism
classes of bimodules, composed through the bimodule tensor product.
The isomorphism class of the canonical bimodule $A\rac A\lac A$ acts
as the unit arrow for the object $A$. Morita equivalence of algebras,
initially defined through so-called equivalence bimodules, is nothing
but isomorphism of objects in \Alg.

We describe an analogous picture for $C^*$- and von Neumann algebras,
algebraic, measured, Lie, and symplectic groupoids, and finally for
Poisson manifolds.  This hinges on the correct identification of the
objects, bimodules, tensor product, and unit arrows. In 
\Ca\ one has \ca s, Hilbert ($C^*$) bimodules, Rieffel's
interior tensor product, and the canonical Hilbert bimodules over \ca s,
in \Wa\ one has \vna s, correspondences, Connes's relative tensor
product, and the standard forms, in \Grb\ (\MG) one has
(measured) groupoids, (measured) functors, composition, and the
identity functors, and in \LG\ (\SyG) one has (symplectic) Lie
groupoids, (symplectic) principal bibundles, the bibundle tensor
product, and the canonical bibundles of  groupoid over
themselves. Finally, in \Po\ one has integrable Poisson manifolds, dual
pairs, symplectic reduction, and symplectic groupoids. In
all cases, known definitions of Morita equivalence turn out to be the
same as isomorphism of objects in the pertinent category.

The Marsden--Weinstein--Meyer reduction procedure in symplectic geometry \cite{AM}
is easily reformulated in terms of the classical tensor product of appropriate
dual pairs; if in the usual formulation one reduces with respect to a group $H$,
one now takes the classical tensor product over the Poisson manifold $\h^*$,
equipped with the Lie--Poisson structure. From this reformulation it is clear
that the reduction procedure
makes sense also for \ca s and \vna s; one merely replaces the category
\Po\ in which classical reduction takes place by the category \Ca\ or \Wa,
and substitutes the group \ca\ $C^*(H)$ or the group \vna\ $W^*(H)$ for
the Poisson manifold $\h^*$.
Reduction is just the composition of certain arrows in
appropriate categories. For operator algebras such a reduction procedure is
automatically regular; the possibility of having singular symplectic quotients
is reserved for the classical setting.

{\bf Acknowledgements} The author is indebted to the participants of
the Seminar on Groupoids 1999--2000, notably M. Crainic, K. Mackenzie,
I. Moerdijk, J. Mr\v{c}un, H. Posthuma, J. Renault, and P. van der Laan for
discussions, and to P. Muhly, J. Schweizer, P. Xu and A. Weinstein for correspondence.

{\bf Notation}
Our notation for categories (including groupoids) will be that $C$
denotes a category as a whole, whose class of objects is $C_0$, and
whose class of arrows (morphisms) is $C_1$. For $a,b\in C_0$, the Hom-space
$(a,b)\subset C_1$ stands for the collection of arrows from $a$ to
$b$.  A functor $F:C\raw D$ decomposes as $F_0:C_0\raw
D_0$ and $F_1:C_1\raw D_1$, subject to the usual axioms.
The unit arrow associated to an object $a\in C_0$ is denoted $1_a\in (a,a)$.
We will only use very elementary aspects of category theory; see
\cite{MacLane} for unexplained definitions.
\section{Algebras}
The original setting for modules, bimodules, tensor products, and
Morita equivalence was pure algebra without additional structure. We
will quickly review this situation; see, e.g.,
\cite{faith}. In the purely algebraic context, all
 algebras under consideration are defined over a fixed commutative
 ring $k$, and are supposed to have a unit.
Note that results for rings may be derived from those for algebras
by regarding a ring as an algebra over $\mathbb{Z}$.
\subsection{The category \Alg\ of algebras
and bimodules}
Recall that an $A$-$B$ bimodule $M$,
written as $A\rac M\lac B$, is a
left $A$ module, which is simultaneously a right $B$ module, such that
both actions commute.  Each algebra $A$ canonically defines 
a bimodule 
\begin{equation}
1_A=A\rac A\lac A,\label{def1A}
\end{equation}
 with actions given by multiplication.
The bimodule tensor product $M\ot_B N$ between
an $A$-$B$ bimodule $M$ and a $B$-$C$ bimodule $N$ is an $A$-$C$
bimodule.  Two $A$-$B$ bimodules $M,M'$ are said to be isomorphic when
there exists an isomorphism $M\raw M'$ as $k$-modules that intertwines
the actions of $A$ and $B$. The bimodule tensor product is merely
associative up to isomorphism, which partly explains the following
definition.
\begin{definition}\label{defAlg}
The category \Alg\ has $k$-algebras as objects,
and isomorphism class\-es of bimodules as arrows. The arrows are composed
by the bimodule tensor product, for which the canonical bimodules
$A$ are   units.
\end{definition}

More precisely, the space of arrows $(A,B)$ from $A$ to $B$ consists
of all isomorphism classes of $A$-$B$ bimodules, the product of two
bimodules $M\in (A,B)$ and $N\in (B,C)$ is $N\circ M=M\ot_B N\in (A,C)$, and
the unit element of $(A,A)$ is the isomorphism class $[1_A]$ of the
bimodule $A\rac A\lac A$.

Here it should be mentioned that $\ot_B$ passes to isomorphism classes.
Whenever no confusion can arise, we will not explicitly mention that 
an arrow is really the isomorphism class of a bimodule, and we will
write $M$ rather than its isomorphism class $[M]$ for such an arrow.
\begin{remark}\label{ringmor}
A bimodule may be regarded as a generalization of a homomorphism;
for given a (unital) homomorphism $\rh:A\raw B$, one constructs a
bimodule $A\rac B\lac B$ by $a\cdot b=\rh(a)b$ and $b\cdot c=bc$.
We write this bimodule as $A\stackrel{\rh}{\rac} B\lac B$.
Note that, with $B=A$, one has $A\stackrel{\id}{\rac} A\lac A=1_A$.

 Let $\rh:A\raw
B$ and $\sg:B\raw C$ be unital homomorphisms, with corresponding
bimodules $A\stackrel{\rh}{\rac} B\lac B$ and $B\stackrel{\sg}{\rac}
C\lac C$.  Denote their tensor product by $\ot^{\sg}_B$.  One then has
the isomorphism $$ A\stackrel{\rh}{\rac} B\ot^{\sg}_B C\lac C\:\:\: \simeq\:\:\:
A\stackrel{\sg\rh}{\rac} C\lac C. 
$$

Hence one obtains a  functor from 
the category of algebras with homomorphisms as arrows into \Alg.
\end{remark}
\subsection{Morita equivalence for algebras}
Morita's theorems give a necessary and sufficient condition for the
representation categories of two algebras to be equivalent.
\begin{definition}\label{Minvt}
An equivalence bimodule  is a bimodule  $A\rac M\lac B$ for which:
\begin{enumerate}
\item $A\simeq \End_{B^{\mbox{\tiny op}}}(M)$;
\item $M$ is finitely generated projective as an $A$- and as a
$B^{\mbox{\tiny op}}$-module.
\end{enumerate}
Two algebras that are related by an equivalence bimodule are called \Me.
\end{definition}

The argument that Morita equivalence defines an equivalence relation
is the same for all cases considered in this paper, so it suffices to
state it here for the special case of algebras. Reflexivity follows from
the existence of the unit bimodules $1_A$.
Symmetry follows from the possibility of turning an equivalence
bimodule $A\rac M\lac B$ around to another equivalence module
$B\rac\ovl{M}\lac A$. In the present case, one has
$\ovl{M}=\Hom_{B^{\mbox{\tiny op}}}(M,B)$, which is a $B$-$A$ bimodule in the obvious
way. Finally, associativity is proved using the bimodule tensor
product: equivalence bimodules $A\rac M_1\lac B$ and $B\rac M_2\lac T$
may be composed to form an equivalence bimodule $A\rac M_1\ot_B
M_2\lac T$.

The categorical interpretation of Definition \ref{Minvt} is as follows.
\begin{proposition}\label{algmor}
An $A$-$B$  bimodule  $M$ is an equivalence bimodule iff its isomorphism
class $[M]\in (A,B)$ is invertible as an arrow in \Alg.

In other words, two algebras are Morita equivalent iff they
are isomorphic objects in \Alg.
\end{proposition}

We write isomorphism of objects in a category as $\cong$.
By definition \cite{MacLane}, one has $A\cong B$ when 
there exist an $A$-$B$ module $M$ and a $B$-$A$ bimodule $M\inv$
such that
\begin{eqnarray}
B\rac M\inv \ot_A M\lac B & \simeq  & B\rac B\lac B;  \label{M1}\\
A\rac M\ot_B M\inv \lac A & \simeq & A\rac A\lac A. \label{M2}
\end{eqnarray}

\begin{Proof}
The  ``$\Law$'' claim is part of ``Morita I'', cf.\ no.\ 12.10.4 
in \cite{faith} for condition 1, and 12.10.2  for condition 2
in Definition \ref{Minvt}. The converse
follows from nos.\  12.8(c) and 4.3(c) in \cite{faith}.
The inverse is
$M\inv=\ovl{M}$, as defined above.
\end{Proof} 

Of course, the fact that Morita
equivalence is an equivalence relation may be rederived from this result.
We now relate Morita equivalence to \rep\ theory. 
\begin{definition}\label{kejo}
The representation category $\Rep(A)$ of an algebra $A$ has left
$A$-modules as objects, and $A$-module maps as arrows.
\end{definition}
The basic statement of Morita theory, then, is as follows.
\begin{proposition}\label{morita}
Two algebras are \Me\ iff they have
equivalent \rep\ categories (where the equivalence functor is required
to be additive). 
\end{proposition}
\begin{Proof}
The idea of the proof in the ``$\Raw$'' direction is as follows.
One first constructs a functor $F:\Rep(B)\raw\Rep(A)$ by taking tensor
products: on \rep s one has $F_0(L)=M\ot_B L\in\Rep(A)_0$ for $L\in
\Rep(B)_0$, and on intertwiners one puts, in obvious notation,
$F_1(f)={\rm id}\ot_B f$. Secondly, one goes in the opposite direction
using $\ovl{M}$, so that one may define a functor
$G:\Rep(A)\raw\Rep(B)$ by means of $G_0(N)=\ovl{M}\ot_A N$, etc.
Proposition \ref{Minverse} then implies that
\begeq
A\rac M\ot_B \ovl{M}\lac A \:\: \cong\:\:  A\rac A\lac A, \label{Minverse}
\end{equation}
where the $\cong$ symbol denotes isomorphism both as a left
and as a right $A$ module. Using this, along with (\ref{def1A}),
one easily shows that $F_0G_0(L)\cong L$ for each $L\in\Rep(A)_0$.
The following property of $F$
is immediate from its definition:
if we denote the intertwiner establishing the isomorphism
$F_0G_0(L)\cong L$ by $\ps_L\in (L,F_0G_0(L))$, then
for each $f\in(K,L)$ one has $f\ps_L=\ps_K F_1G_1(f)$.  
Thus one has constructed an equivalence functor $F$.

In the ``$\Law$'' direction, one constructs $M$, given an equivalence
functor $F:\Rep(B)\raw\Rep(A)$, by putting $M=F_0(1_B)$, where
initially $B$ is seen as a left $B$ module.  The left $B$ action on
$B$ is turned into a left $A$ action on $M$ by definition,
but in addition the right $B$ action on $B$ is turned into a right $B$
action on $M$ through $F_1$, since $B\op\subseteq (B,B)\subset
\Rep(B)_1$. The definition of an equivalence functor then 
implies that $M$ is invertible in \Alg. 
\end{Proof}

The first part of this proof trivially generalizes to all other
classes of mathematical objects we study in this paper. The second
part, on the other hand, only generalizes when the analogues of the
identity intertwiners $1_A$ lie in the \rep\ category under consideration,
and when there is enough intertwining around to turn the analogues of
$F_0(1_B)$ into an invertible bimodule. 
\section{Operator algebras}
A technical reference for the theory of \ca s and \vna s is \cite{KR1,KR2}.
For an unsurpassed overview, see \cite{Con}.
\subsection{The category \Ca\ of \ca s and Hilbert bimodules}
It may not be all that obvious, but the correct $C^*$-algebraic
analogue of a bimodule for algebras is a so-called Hilbert bimodule.
First, recall the concept of a Hilbert module (alternatively called a
$C^*$-module \ci{Con} or a Hilbert $C^*$-module \ci{Lance,RW,NPL3})
over a given \ca\ $\B$, due to Paschke \cite{Pas} and Rieffel \cite{Rie1,Rie2}.
\begin{definition}\label{defhb}
 A Hilbert module over a \ca\ $\B$ is a complex linear space $\CE$ equipped
with a right action of $\B$ on $\CE$ and a compatible $\B$-valued
inner product, that is, a sesquilinear map $\langle \,
,\,\ra_{\B}:\CE\x\CE\raw\B$, linear in the second and antilinear in
the first entry, satisfying $\langle \Ps,\Ph\ra_{\B}^* =
\langle\Ph,\Ps\ra_{\B}$. One requires $\langle\Ps,\Ps\ra_{\B} \geq 0$,
and $\langle\Ps,\Ps\ra_{\B} = 0$ iff $\Ps=0$.  The space $\CE$ has to
be complete in the norm 
\begeq
\n\Ps\n^2= \n\langle\Ps,\Ps\ra_{\B}\n. \label{hcnorm}
\end{equation} The
compatibility condition is 
\begeq
\langle\Ps,\Ph B\ra_{\B} =
\langle\Ps,\Ph\ra_{\B} B. \label{comp}
\end{equation}
\end{definition}

For example, a Hilbert space is a Hilbert module over $\C$.

A map $A:\CE\raw\CE$ for which there exists $A^*:\CE\raw\CE$
such that $\la \Ps,A\Ph\ra_{\B}=\la A^*\Ps,\Ph\ra_{\B}$ is called
adjointable.  An adjointable map is automatically $\C$-linear,
$\B$-linear, and bounded.  The adjoint of an adjointable map is
unique, and the map $A\mapsto A^*$ defines an involution on the space
$\CL_{\B}(\CE)$ of all adjointable maps on $\CE$. This space thereby
becomes a \ca.
\begin{definition}\label{defhbm}
An $\A$-$\B$ Hilbert bimodule, where $\A$ and $\B$ are \ca s, is 
 a Hilbert module $\CE$ over $\B$, along with a nondegenerate
$\mbox{}^*$-homomorphism of $\A$ into $\CL_{\B}(\CE)$. We write
$\A\rac\CE\rlh\B$.
\end{definition}

This concept is due to Rieffel \cite{Rie1,Rie2}, who originally spoke
of Hermitian $\B$-rigged $\A$-modules rather than $\A$-$\B$ Hilbert
bimodules (one sometimes also calls $\CE$ a $C^*$-correspondence
between $\A$ and $\B$; cf.\ \cite{BDH,MS}), and did not impose the
nondegeneracy condition.  The latter means that $\A\CE$ be dense in
$\CE$ \cite{Lance}; when $\A$ is unital, it obviously suffices that
the $\mbox{}^*$-homomorphism preserves the unit.  In other words, one
has a space with a $\B$-valued inner product and compatible left $\A$
and right $\B$-actions, where it should be remarked that the left and
right compatibility conditions are quite different from each other.
Note that an $\A$-$\B$ Hilbert bimodule is an algebraic $\A$-$\B$
bimodule, since $\CL_{\B}(\CE)$ commutes with the right $\B$-action.

For example, a Hilbert space is a $\C$-$\C$  Hilbert bimodule
in the obvious way, as well as a $\BH$-$\C$ Hilbert bimodule, or an
$\A$-$\C$ Hilbert bimodule, where $\A\subset\BH$ is some \ca. 
The following example is the \ca ic version of the canonical algebra bimodule
$A\rac A\lac A$.
\begin{example} \label{BB} Each \ca\
defines a Hilbert bimodule $\B\rac\B\rlh\B$ over itself, in which
$\langle A,B\ra_{\B} =A^*B$, and the left and right actions are given
by left and right multiplication, respectively. This Hilbert bimodule
will be called  $1_{\B}$.
\end{example}

  Note that the
$C^*$-norm in $\B$ coincides with its norm as a Hilbert module because
of the $C^*$-axiom $\n A^*A\n=\n A\n^2$.  

We turn to the \ca ic analogue of the bimodule tensor product; given
an $\A$-$\B$ Hilbert bimodule $\CE$ and a $\B$-$\GC$ Hilbert bimodule
$\CF$, we wish to define an $\A$-$\GC$ Hilbert bimodule
$\CE\hat{\ot}_{\B}\CF$. The explicit construction of this ``interior''
tensor product, due to Rieffel \cite{Rie1,Rie2}, is as follows.

One first defines a $\GC$-valued inner product on the algebraic
tensor product $\CE\ot_{\C}\CF$ by sesquilinear extension of 
\begeq \la
\Ps_1\ot\Ps_2,\Ph_1\ot\Ph_2\ra^{\ot}_{\GC}=
\la\Ps_2,\la\Ps_1,\Ph_1\ra_{\B}\Ph_2\ra_{\GC}. \ll{PsBcPs} 
\end{equation} This is
positive semidefinite, and combined with the norm on $\GC$ one obtains
a seminorm on $\CE\ot_{\C}\CF$, as in (\ref{hcnorm}).  The completion
of the quotient of $\CE\ot_{\C}\CF$ by the null space of $\la\,
,\,\ra^{\ot}_{\GC}$ is $\CE\hat{\ot}_{\B}\CF$ as a vector space. The
crucial point is that $\CE\hat{\ot}_{\B}\CF$ inherits the left action
of $\A$ on $\CE$, the right action of $\GC$ on $\CF$, and also the
$\GC$-valued inner product (\ref{PsBcPs}), so that
$\CE\hat{\ot}_{\B}\CF$ itself becomes an $\A$-$\GC$ Hilbert
bimodule. Many good features of  Rieffel's interior tensor product are caused
by the fact that the null space in question is precisely the closed
linear span of all expressions of the form $\Ps
B\ot_{\C}\Ph-\Ps\ot_{\C} B\Ph$; cf.\ \cite{Lance}.
One sometimes denotes the image (projection) of a vector
$\Ps\ot\Ph\in\CE\ot_{\C}\CF$ in $\CE\hat{\ot}_{\B}\CF$ by $\Ps\ot_{\B}\Ph$.

The canonical bimodule $1_{\B}$ of Example \ref{BB} is a unit for
Rieffel's tensor product $\hat{\ot}_{\B}$. Thus we obrain a \ca ic
version of Definition \ref{defAlg}, in which two $\A$-$\B$ Hilbert
bimodules $\CE,\CF$ are called isomorphic when there is a unitary
$U\in\CL_{\B}(\CE,\CF)$; cf.\ \cite{Lance}, p.\ 24.
\begin{definition}\label{defCa}
The category \Ca\ has $C^*$-algebras as objects, and isomorphism
class\-es of Hilbert bimodules as arrows. The arrows are composed by
Rieffel's  interior tensor product, for which the canonical Hilbert
bimodules $1_{\A}$ are   units.
\end{definition}

This category was introduced independently in \cite{Sch},
and, in the guise of a bicategory (where the arrows are Hilbert
bimodules rather than isomorphism classes thereof), in
\cite{NPLDR}.  
Along the lines of Remark \ref{ringmor}, we have
\begin{remark}\label{cstamor}
Given a nondegenerate $\mbox{}^*$-homomorphism $\rh:\A\raw\B$, 
one constructs an  $\A$-$\B$ Hilbert bimodule $\A\rac\B\rlh\B$ by
$A(B)=\rh(A)B$, and the other operations as in Example \ref{BB}.
(Here $\rh$ is nondegenerate when $\rh(\A)\B$ is dense in $\B$.)
We write $\A\stackrel{\rh}{\rac} \B\rlh\B$.
Thus one obtains a  functor from the category of
$C^*$-algebras with $\mbox{}^*$-homomorphisms as arrows into \Ca.
\end{remark}
\subsection{Morita equivalence for \ca s}
The \ca ic version of Definition \ref{Minvt}, due to Rieffel \cite{Rie1,Rie2} is as follows.
\begin{definition}\label{smeca}
A Hilbert bimodule  $\CM\in (\A,\B)$ is called an equivalence
Hilbert bimodule when:
\begin{enumerate}
\item the linear span of the range of $\la \, ,\,\bb$ is dense in $\B$
(in other words, $\CM\rlh\B$ is full);
\item
the $\mbox{}^*$-homomorphism of $\A$ into $\CL_{\B}(\CE)$ of
Definition \ref{defhbm} is an isomorphism $\A\simeq\CK_{\B}(\CM)$.
(If $\A$ has a unit, this isomorphism will be $\A\simeq\CL_{\B}(\CM)$.)
\end{enumerate}
Two \ca s that are related by an equivalence Hilbert bimodule are called
\Me.
\end{definition}

Here $\CK_{\B}(\CE)$ is the \ca\ of ``compact'' operators on a Hilbert
module $\CE$ over a \ca\ $\B$ \cite{Rie1,Rie2,Lance,RW,NPL3}.  This is
the norm-closed algebra generated by all operators on $\CM$ of the
type $\theta_{\Ps,\Ph}Z=\Ps\la\Ph,Z\ra_{\B}$. 

It can be shown that two unital
\ca s are Morita equivalent as \ca s iff they are  Morita equivalent as algebras \cite{Bee}.
Another nontrivial result is that two $\sg$-unital \ca s (i.e., having
a countable approximate unit) are Morita equivalent iff they are
stably isomorphic (in that they become $\mbox{}^*$-isomorphic after
tensoring with the
\ca\ of compact operators on some \Hs). See \cite{BGR,Lance}.

A number of equivalent conditions for Morita equivalence of \ca s are
given in \cite{RW}, which is a good reference for the subject. Note
that Rieffel, and many later authors, use the term ``strongly \Me'' to
describe the situation in Definition \ref{smeca}. As in Proposition
\ref{algmor}, we have
\begin{proposition}\label{calgmor}
A Hilbert bimodule $\A\rac\CM\rlh\B$ is an equivalence Hilbert bimodule
iff its isomorphism class $[\CM]\in (\A,\B)$
is invertible as an arrow in \Ca.

In other words, two $C^*$-algebras are Morita equivalent iff they are
isomorphic objects in \Ca.
\end{proposition}

This result was conjectured by the author in the setting of
bicategories (cf.\ \cite{NPLDR}), after which Paul Muhly pointed out
that the difficult half (``$\Law$'') of the proof of the categorical
version of the claim, as formulated above, had already been given by
Schweizer (see Prop.\ 2.3 in \cite{Sch}).  This part of the proof
below is a rearrangement of Schweizer's proof, with considerable
detail added.

\begin{Proof} 
The ``$\Raw$'' claim is a rephrasing of the statement that, given the
conditions in Definition
\ref{smeca}, an inverse $\CM\inv$ of $\CM$ exists, in satisfying
the \ca ic analogue of (\ref{M1}) and (\ref{M2}), viz.
\begin{eqnarray}
\B\rac \CM\inv \otA \CM\rlh \B & \simeq  & \B\rac \B\rlh \B;  \label{CM1}\\
\A\rac \CM\otB \CM\inv \rlh \A & \simeq & \A\rac \A\rlh \A. \label{CM2}
\end{eqnarray}

This was stated (without proof) by Rieffel
on p.\ 239 of \cite{Rie1}, and is proved  in \cite{RW}, Prop.\ 3.28.
For later use, we here merely recall that the inverse Hilbert bimodule
is $\CM\inv=\ovl{\CM}$,  the conjugate space of
$\CM$, on which $\B$ acts from the left by $B:\Ps\mapsto \Ps B^*$, and
$\A$ acts from the right by $A:\Ps\mapsto A^*\Ps$. The $\A$-valued
inner product on $\ovl{\CM}$ is given by $\la\Ps,\Ph\ra_{\A}
=\phv\inv(\theta_{\Ps,\Ph})$.  

For the ``$\Law$'' direction, assume the existence of $\CM\inv$ such
that (\ref{CM1}) and (\ref{CM2}) hold.  First, note that condition 1 in
Definition \ref{smeca} trivially follows from (\ref{CM1}) and the
definition of the $\B$-valued inner product on $\CM\inv\otA\CM$.

 We denote the map $\B\rightarrow\CL_{\A}(\CM\inv)$ by $\rh$.  Let
 $\hat{\rh}_*:\CL_{\B}(\CM)\raw \CL_{\A}(\A)$ be the composition of the
 canonical map $\rh_*:\CL_{\B}(\CM)\raw \CL_{\A}(\CM\otB\CM\inv)$ 
(cf.\ \cite{Lance}, p.\ 42) with
 the isomorphism $\CL_{\A}(\CM\otB\CM\inv)\simeq \CL_{\A}(\A)$ given by
 (\ref{CM2}).  Recall that for the canonical Hilbert $C^*$-module
 $\A\rlh\A$ one has $\CL_{\A}(\A)=M(\A)$, the multiplier algebra of
 $\A$
\cite{Lance,RW}, so that $\hat{\rh}_*:\CL_{\B}(\CM)\raw M(\A)$.
Also, $\CK_{\A}(\A)=\A$, which is the first $\A$ on the right-hand side of
(\ref{CM2}), regarded as a subalgebra of $M(\A)$.
Note, then, that by construction one has
\begeq
\hat{\rh}_*\circ\phv(A)=A \label{bycon}
\end{equation}
 for all $A\in\A$.  It follows from (\ref{CM1}) that $\rh$ is
 injective, so that $\rh_*$ is injective (cf.\ \cite{Lance}, p.\
 42), and hence $\hat{\rh}_*$ is injective.
We now claim that
\begeq
\hat{\rh}_*(\CK_{\B}(\CM))\subseteq\A, \label{step1}
\end{equation}
where $\A$ is seen as a subalgebra of $M(\A)$.
For, since $\A$ is an ideal in $M(\A)$, we have $E_i \hat{\rh}_*(K)\in
\A$ for all $K\in\CL_{\B}(\CM)$ (where $\{E_i\}$ is an approximate unit
in $\A$).  Using (\ref{bycon}), one has $E_i
\hat{\rh}_*(K)=\hat{\rh}_*(\phv(E_i)K)$.  Hence, since $\hat{\rh}_*$
is contractive, and the norm induced on $\A$ by its embedding in
$M(\A)$ is the original norm, one has $$
\| E_i \hat{\rh}_*(K) -\hat{\rh}_*(K)\|_{\A} \leq
\| \phv(E_i) K -K\|.
$$ 
Now, for $K\in\CK_{\B}(\CM)$ one has $E_i \hat{\rh}_*(K) \raw\hat{\rh}_*(K)$
by Lemma \ref{slem} below, from which  (\ref{step1}) follows.

Similarly,  exchanging $\A$ and $\B$ and $\CM$ and
$\CM\inv$, one obtains
\begeq
\hat{\phv}_*(\CK_{\A}(\CM\inv))\subseteq\B, \label{step2}
\end{equation}
where  $\hat{\phv}_*:\CL_{\A}(\CM\inv)\raw M(\B)$ is 
$\phv_*:\CL_{\A}(\CM\inv)\raw \CL_{\B}(\CM\inv\ot_{\A}\CM)$
composed with the isomorphism $\CL_{\B}(\CM\inv\ot_{\A}\CM)\simeq M(\B)$
induced by (\ref{CM1}).
The  inclusion (\ref{step2}) will now be used to show that 
\begeq
\phv(\A)\subseteq \CK_{\B}(\CM).\label{step3}
\end{equation}
Using the isomorphism $\A\simeq \ovl{\CM\inv}\otq_{\CK_{\A}(\CM\inv)}\CM\inv$
as $\A$-$\A$ Hilbert bimodules (cf.\ the ``$\Raw$'' part of the proof),
the fact that $\A$ is a unit for $\ot_{\A}$,  the associativity of
Rieffel's tensor product up to isomorphism, and (\ref{CM1}),
one obtains
\begin{eqnarray}
\CM &\simeq & \A\otA\CM\simeq (\ovl{\CM\inv}\otq_{\CK_{\A}(\CM\inv)}\CM\inv)
\otA\CM \nn \\
& \simeq & \ovl{\CM\inv}\otq_{\CK_{\A}(\CM\inv)}(\CM\inv
\otA\CM) \simeq \ovl{\CM\inv}\otq_{\CK_{\A}(\CM\inv)}\B \label{boiso}
\end{eqnarray}
as $\A$-$\B$ Hilbert bimodules. 

It follows from (\ref{CM2}) and the
definition of the $\A$-valued inner product on $\CM\otB\CM\inv$
that $\CM\inv\rlh\A$ is full. For any full Hilbert $C^*$-module
$\CE\rlh\A$, one has an isomorphism of \ca s
\begeq
\A\simeq \CK_{\CK_{\A}(\CE)}(\ovl{\CE}); \label{keq}
\end{equation}
this is simply the proof of symmetry in the standard argument that
(strong) Morita equivalence is indeed an equivalence relation \cite{Rie1,RW}
(or see Thm.\ IV.2.3.3 in \cite{NPL3} for a direct proof of (\ref{keq})).
Hence $\A\simeq \CK_{\CK_{\A}(\CM\inv)}(\ovl{\CM\inv})$. Combining this
with Prop.\ 4.7 in \cite{Lance}, which applies because of
(\ref{step2}) and the fact that $\B\simeq\CK_{\B}(\B)$,
 yields 
\begeq
\A\subseteq
\CK_{\B}(\ovl{\CM\inv}\otq_{\CK_{\A}(\CM\inv)}\B),\label{88}
\end{equation}
where we have suppressed the notation for a number of isomorphisms and
other maps. Combining (\ref{88}) and (\ref{boiso})  gives (\ref{step3}).

It now follows from (\ref{step1}), (\ref{step3}), and (\ref{bycon})
that  $\hat{\rh}_*:\CK_{\B}(\CM)\raw\A$ is surjective. Since we already
know that it is injective, $\hat{\rh}_*$ defines an isomorphism from
$\CK_{\B}(\CM)$ to $\A$. 
Eq.\ (\ref{bycon})  then shows that $\phv:\A\raw
\CK_{\B}(\CM)$ is the inverse of $\hat{\rh}_*$, so that it must be
an isomorphism as well.
\end{Proof} 

In this proof, we used the following lemma.
\begin{lemma}\label{slem}
Let $\A\rac\CE\rlh\B$ be a Hilbert bimodule, and let $\{E_i\}$ be
an approximate unit in $\A$. Then for each $K\in \CK_{\B}(\CE)$
one has
\begin{equation}
\lim_i \| \phv (E_i)K-K\| =0 \label{limi}
\end{equation}
in the usual norm on $\CL_{\B}(\CE)$.
\end{lemma}
\begin{Proof} 
Prop.\ 2.5(iii) in \cite{Lance} shows that the nondegeneracy of $\phv$
implies that $\| \phv (E_i)\Ps-\Ps\|\raw 0$ for each $\Ps\in\CE$.
Using the elementary bounds $\|\Ps B\|\leq \| B\|\, \|\Ps\|$ and
$\|\la\Ph,Z\ra\|\leq \|\Ph\|\, \| Z\|$, it easily follows that
$\| \phv (E_i)\theta_{\Ps,\Ph}-\theta_{\Ps,\Ph}\|\raw 0$
for all $\Ps,\Ph\in\CE$. Using the
definition of $\CK_{\B}(\CE)$ and the uniform bound $\| E_i\|\leq 1$,
eq.\ (\ref{limi}) follows.
\end{Proof} 

Although a \ca\ is an algebra over $\mathbb{C}$
(though not always a unital one), 
one seeks \rep s on Hilbert spaces rather than 
on general complex vector spaces.
Thus Rieffel, who launched the theory of Morita
equivalence of \ca s \cite{Rie2}, defined the representation category
$\Rep(\A)$ of a \ca\ $\A$ as follows.
\begin{definition}\label{repsca}
The \rep\ category $\Rep(\A)$ of a \ca\ $\A$ has nondegenerate
$\mbox{}^*$-\rep s of $\A$ on a Hilbert space as objects, and bounded
linear intertwiners as arrows.
\end{definition}
Taking the polar decomposition of an invertible arrow, it follows
that isomorphism in this category is unitary equivalence, as it should.
The adaptation of  Proposition \ref{morita} to \ca s, again due
to Rieffel, now reads
\begin{proposition}\label{rit}
If two \ca s are \Me, then they have equivalent \rep\ categories
(where the equivalence functor is required to be linear and
$\mbox{}^*$-preserving on intertwiners).
\end{proposition}

The proof is the same as for algebras, with the obvious replacements.
It is clear that the purely algebraic  proof of a potential ``$\Law$''
part of Proposition \ref{rit} cannot immediately be adapted to the present
case, since the bimodule $1_{\A}$ is not itself an element of
$\Rep(\A)$. Indeed, Rieffel's Morita theorem for \ca s only has the
``$\Raw$'' implication.
To remedy this defect, one should enlarge $\Rep(\A)$ so that it
contains $\A$.  This has been done by Blecher \cite{Ble2} in the
setting of operator spaces, operator modules, and completely bounded
maps.
\subsection{The category \Wa\ of \vna s and correspondences}
A \vna\ $\GM$ is a unital \ca\ that is the dual of a Banach space,
 its so-called predual $\GM_*$. Hence, in addition to its norm topology,
it comes equipped with a second natural topology, namely the pertinent
weak$\mbox{}^*$, or $\sg(\GM,\GM_*)$, or $\sg$-weak topology.
Maps between \vna s are always required to be $\sg$-weakly continuous;
such maps are said to be normal.
For example, normality of a unital \rep\ $\pi$ guarantees that $\pi(\GM)$ is
$\sg(\BH,\BH_*)$-closed, which, by von Neumann's bicommutant theorem,
is equivalent to the property $\pi(\GM)''=\pi(\GM)$. 

Although one may adapt the theory of Hilbert bimodules for \ca s so as
to include normality of the actions, as in \cite{Rie2}, there is a
much simpler approach to bimodules for \vna s, initiated by Connes
\cite{Con}.
\begin{definition}\label{defcor}
Let $\GM,\GN$ be \vna s. An $\GM$-$\GN$ correspondence
$\GM\rac\H\lac\GN$ is given by a \Hs\ $\H$ with normal unital \rep s
$\pi(\GM) $   and $\phv(\GN\op)$ on $\H$,  such that 
$\pi(\GM)\subseteq \phv(\GN\op)'$ (and hence $\phv(\GN\op)\subseteq
\pi(\GM)'$).
\end{definition}

In what follows, we usually omit the symbols $\pi$ and $\phv$, implying
that these \rep s are injective.
The notion of isomorphism of correspondences is the obvious one:
one requires a unitary isomorphism between the Hilbert spaces
in question that intertwines the left and right actions.

The following structure will be important \cite{Con,KR2}.
\begin{definition}\label{defsf}
A \vna\ $\GM$ acting on a \Hs\ $\H$ is said to be in standard form
when there is a conjugation $J$ on $\H$ (that is, an antiunitary
operator squaring to $\I$) such that $\GM'\simeq\GM\op$
through the (linear) map $A\mapsto JA^*J$, under which the center
of $\GM$ is pointwise invariant.
\end{definition}

Hence a standard form defines an  $\GM$-$\GM$ correspondence
$\GM\rac\H\lac\GM$ with the property that $\GM'\simeq\GM\op$. A
corollary of the Tomita--Takesaki theory \cite{Con,KR2} is
\begin{remark}\label{TT}
Each \vna\ $\GM$ is isomorphic to one in standard form, and the standard
form is unique up to unitary equivalence. 
\end{remark}

We write ``the'' standard form of $\GM$ as $\GM\rac L^2(\GM)\lac\GM$,
where the symbol $L^2(\GM)$ for the \Hs\ in question is purely
notational, and has nothing to do with $L^2$ functions on $\GM$. In
the theory of finite \vna s it is the completion of $\GM$ with respect
to the inner product given by the normalized trace, which indeed
yields the structure in Remark \ref{TT}. The simplest example is
$\GM=M_n(\C)$, for which one may take $L^2(\GM)=\GM$, with inner
product $(M,N)=(1/n)\Tr M^*N$ and obvious left and right actions. For
$\Om$ one may take the unit matrix.  For general \vna s a canonical
construction of the standard form is given in \cite{Con}, App.\ V.B.

The correspondence $\GM\rac L^2(\GM)\lac\GM$ is the von
Neumann-algebraic counterpart of the canonical bimodule $1_A$ for
algebras, and of Example \ref{BB} for \ca s; its isomorphism class
 will play the role of the unit arrow at
$\GM$ in the category \Wa\ to be defined.  In addition, it plays a
central role in the construction of the tensor product $\GM\rac
\H\boxtimes_{\GN}\CK\lac\GP$ of an $\GM$-$\GN$ correspondence with an
$\GN$-$\GP$ correspondence \cite{Con}, which we now review, following
\cite{Sau}.

For simplicity, we assume that
$\GN$ is $\sg$-finite (that is, every family of mutually orthogonal
projections is at most countable; this is true, for example, when
$\GN$ acts on a separable
\Hs, or has separable predual). This implies that $L^2(\GN)$
contains a unit vector $\Om$ such that $\GN\Om$ and $\Om\GN$ are dense
in $L^2(\GN)$.  Using the theory of weights \cite{KR2}, all
constructions below may be modified so as to apply to the general
case. The dependence on $\Om$ is immaterial, up to isomorphism of
correspondences.  The following lemma,
due to Connes \cite{Con0}, is crucial.
\begin{lemma}\label{saulem}
Let $\GM\rac\H\lac\GN$ be a correspondence.
Define $\til{\H}\subset\H$ by the property that for each
$\Ps\in\til{\H}$ the linear map $R_{\Ps}:L^2(\GN)\raw\H$, defined on the
dense domain $J\GN\Om$ by $R_{\Ps}(JA^*\Om)=\Ps A$, is bounded.  
Then:
\begin{enumerate}
\item The subspace $\til{\H}$ is stable under $\GM$ and $\GN$.
\item
The subspace $\til{\H}$ is dense in $\H$.
\item
The relation $\Ps\lraw R_{\Ps}$ is a bijection between $\til{\H}$
and $\Hom_{\GN\op}(L^2(\GN),\H)$.
\item For $\Ps,\Ph\in\til{\H}$ one has $R^*_{\Ps}R_{\Ph}\in\GN$,
 where $\GN$ is identified 
with its (left) \rep\ on $L^2(\GN)$.
\end{enumerate}
\end{lemma}

Now, given a second correspondence $\GN\rac\CK\lac\GP$,
one equips $\til{\H}\ot_{\C}\CK$ with the sesquilinear form defined
by sesquilinear extension of
\begeq
(\Ps\ot\ \ps,\Ph\ot \phv)_0=(\ps,R^*_{\Ps}R_{\Ph}\phv)_{\CK},\label{form}
\end{equation}
which is well defined because of Lemma \ref{saulem}.4. This form is positive
semidefinite, hence a pre-inner product, and the completion of the
quotient of $\til{\H}\ot_{\C}\CK$ by its null space is a \Hs, denoted
by $\H\boxtimes_{\GN}\CK$.  This is alternatively called the relative tensor
product \cite{Sau}, Connes fusion \cite{Was}, or Connes's tensor product 
of $\H$ and
$\CK$ over $\GN$.  The left $\GM$ and right $\GP$
actions quotient to $\H\boxtimes_{\GN}\CK$, yielding a new
correspondence $\GM\rac \H\boxtimes_{\GN}\CK\lac\GP$. 
It is easily verified that this composition is associative up to isomorphism.
The notation $\Ps\boxtimes_{\GN}\phv$ for the image of
$\Ps\ot\phv$ in $\H\boxtimes_{\GN}\CK$ will occasionally be used.

In an equivalent description  that treats $\H$ and
$\CK$ in a more symmetric way, one analogously defines $\til{\CK}\subset\CK$
as the space of vectors $\ps\in\CK$ for which the map
$L_{\ps}:L^2(\GN)\raw\CK$, defined on the dense domain $\GN\Om$ by
$L_{\ps}(A\Om)=A\ps$, is bounded.  Again,  $\til{\CK}$ is
dense in $\CK$, and one has a bijection between
$\til{\CK}$ and $\Hom_{\GN}(L^2(\GN),\CK)$.
 Thus one may initially define the form
(\ref{form}) on $\til{\H}\ot_{\C}\til{\CK}$ by
\begeq
(\Ps\ot\ \ps,\Ph\ot \phv)_0=(L_{\ps}\Om,R^*_{\Ps}R_{\Ph}L_{\phv}\Om)_{\CK}
= (\Om,R^*_{\Ps}R_{\Ph}L_{\ps}^*L_{\phv}\Om)_{L^2(\GN)}
,\label{form2}
\end{equation}
where we have used the fact that $L_{\ps}^*\in
\Hom_{\GN}(\CK,L^2(\GN))$  with Lemma \ref{saulem}.4.  Thus one
may define $(\, ,\,)_0$ on $\Hom_{\GN\op}(L^2(\GN),\H)\ot_{\C}
\Hom_{\GN}(L^2(\GN),\CK)$ from the start by 
\begeq
(A_1\ot B_1,A_2\ot B_2)_0=(\Om, A_1^* A_2 B_1^* B_2\Om)_{L^2(\GN)}
.\label{form3}
\end{equation}
 The $\GM$-action on
$\H$ and the $\GP$ action on $\CK$ may be moved to
$\Hom_{\GN\op}(L^2(\GN),\H)$ and to $\Hom_{\GN}(L^2(\GN),\CK)$,
respectively, in the obvious way, and subsequently descend to the
quotient, as before.  This is the definition of Connes fusion 
used by Wassermann \cite{Was}, who also proves associativity up to
isomorphism.
\begin{lemma}\label{CSlem}
The standard form of a \vna\ $\GM$ (see Remark \ref{TT}) is a unit
for Connes's tensor product $\boxtimes_{\GM}$, up to
isomorphism. 
\end{lemma}

For the proof cf.\ \cite{Sau}, no.\ 2.4.
Hence we have
the \vna ic version of Definitions \ref{defAlg} and \ref{defCa}:
\begin{definition}\label{defWa}
The category \Wa\ has \vna s as objects, and isomorphism classes of
correspondences as arrows, composed by Connes's
relative tensor product, for which the standard forms $L^2(\GM)$ are
units.
\end{definition}

As in Remarks \ref{ringmor} and \ref{cstamor}, one has
\begin{remark}\label{vnamor}
A normal unital $\mbox{}^*$-homomorphism
$\rh:\GM\raw\GN$ defines a correspondence
$\GM\stackrel{\rh}{\rac} L^2(\GN)\lac\GN$ by 
$A(\Ps)=\rh(A)\Ps$ and $(\Ps)B= \Ps B$, where $A\in\GM$, $B\in\GN$,
$\Ps\in L^2(\GN)$.
Thus one obtains a  functor from the category of
\vna s with normal unital $\mbox{}^*$-homomorphisms as arrows into \Wa.
\end{remark}
\subsection{Morita equivalence for \vna s}
The theory of Morita equivalence of \vna s was initiated by
Rieffel \cite{Rie2}. His
 definition  of strong Morita equivalence
was directly adapted from his \ca ic Definition \ref{smeca}.  However,
the theory of correspondences enables one to rewrite his theory in a
way that practically copies the purely algebraic case.
\begin{definition}\label{Minvna}
A  correspondence $\GM\rac\H\lac\GN$ is called an
equivalence correspondence when
$\GM'\simeq\GN\op$.
Two \vna s that are related by an equivalence correspondence are called
\Me.
\end{definition}

Rieffel's original
definition of strong Morita equivalence (Def.\ 7.5 in \cite{Rie2})
 is equivalent to Definition \ref{Minvna} by
Thms.\ 7.9 and 8.15  in \cite{Rie2}. The
\ca ic characterization of Morita equivalence of \cite{BGR}
has an easier \vna ic counterpart, namely  that two
\vna s are Morita equivalent iff they are stably isomorphic
(which this time means that they become $\sg$-weakly $\mbox{}^*$-isomorphic
after tensoring with $\BH$ for some \Hs\ $\H$).

As in Propositions \ref{algmor} and \ref{calgmor}, one finds
\begin{proposition}\label{vnalgmor}
A correspondence $\GM\rac\H\lac\GN$ is an equivalence correspondence
iff its isomorphism class $[\H]\in (\GM,\GN)$
is invertible as an arrow in \Wa.

In other words, two \vna s are Morita equivalent iff they
are isomorphic objects in \Wa.
\end{proposition}
\begin{Proof}
For ``$\Raw$'' see Prop.\ 3.1 in \cite{Sau}.
 The inverse of an invertible
correspondence $\GM\rac\H\lac\GN$, with $\GN\op=\GM'$, is $\GN\rac\ovl{\H}\lac\GM$, defined
as for Hilbert bimodules (see the proof of Proposition
\ref{calgmor}).

 For the converse implication, note that
given an invertible correspondence $\GM\rac\H\lac\GN$, with inverse
$\GN\rac\H\inv\lac\GM$, by assumption one has
\begin{eqnarray}
\GN\rac \H\inv \boxtimes_{\GM}\H \lac \GN & \simeq  & \GN\rac L^2(\GN)\lac \GN;  \label{Mvn1}\\
\GM\rac \H\boxtimes_{\GN}\H\inv  \lac \GM & \simeq & \GM\rac L^2(\GM)\lac \GM. \label{Mvn2}
\end{eqnarray}
Compare (\ref{M1}) and (\ref{M2}). Using Prop.\ 3.3 in \cite{Sau}, eq.\ (\ref{Mvn1}) implies
\begin{equation}
\GN\boxtimes_{\GM}\I_{\H}=(\GM\op)'\boxtimes_{\GM}\I_{\H}, \label{csten}
\end{equation}
where $(\GM\op)'$ is the commutant of $\GM\op$ on $\H\inv$. 
This is an equality of \vna s on the \Hs\ $\H\inv \boxtimes_{\GM}\H$. Now 
use the first part of the proof with $\GN=(\GM')\op$ on $\H$, so that
$\H\boxtimes_{(\GM')\op}\ovl{\H}\simeq L^2(\GM)$ as  $\GM$-$\GM$ correspondences.
Using associativity of the tensor product up to isomorphism, and finally throwing in
Lemma \ref{CSlem}, one obtains
$$
(\H\inv \boxtimes_{\GM}\H)\boxtimes_{(\GM')\op}\ovl{\H}\simeq \H\inv
$$ as $\GN$-$\GM$ correspondences. Hence
(\ref{csten}) implies $\GN=(\GM\op)'$ on $\H\inv$.
Similarly, (\ref{Mvn1})  implies
$\GM=(\GN')\op$ on $\H$.
\end{Proof}
\begin{definition} \label{repvn}
Let $\GM$ be a \vna. The \rep\ category $\Rep(\GM)$ has normal unital
$\mbox{}^*$-\rep s on \Hs s as objects, and boun\-ded linear
intertwiners as arrows.
\end{definition}
\noindent
Unitality is equivalent to nondegeneracy; cf.\ Definition \ref{repsca}.

Rieffel's Morita theorem for \vna s  then reads as follows:
\begin{proposition}\label{MVNA}
Two \vna s are  related by an equivalence correspondence iff their
\rep\ categories are equivalent (and the equivalence functor
implementing $\simeq$ is linear and $\mbox{}^*$-preserving on intertwiners).
\end{proposition}
\begin{Proof}
For Rieffel's own proof cf.\ \cite{Rie2}. A proof that is virtually
the same as for algebras may be based on Connes's relative tensor
product (replacing the bimodule tensor product), and the standard form
of a \vna\ (replacing the canonical algebra bimodule).  Also cf.\
Prop.\ 3.5.1 in \cite{Sau}.  Note that such a proof could be mapped
onto Rieffel's through Proposition \ref{BDH2.2} below.
\end{Proof}
\subsection{The connection between correspondences and 
Hilbert bimodules} \label{cchbm}
Since \vna s are \ca s with additional structure,
one could look at $\GM$-$\GN$ Hilbert bimodules as well as at
$\GM$-$\GN$ correspondences. The precise connection between
correspondences and Hilbert bimodules for \vna s was established in
Thm.\ 2.2 in
\cite{BDH}, as follows.  The following terminology is used.  A
Hilbert module $\CE\rlh\B$ is selfdual when any bounded $\B$-linear
map $A:\CE\raw\B$ is of the form $A(\Ps)=\la \Ph,\Ps\bb$ for some
$\Ph\in\CE$.  It is normal when all maps $A\mapsto
\la\Ps,A\Ph \ra_{\GN}$ from $\GM$ to $\GN$ are normal.
\begin{proposition}\label{BDH2.2}
\begin{enumerate}
\item
Let $\GM\rac\H\lac\GN$ be a correspondence. Then 
$$
\GM\rac \Hom_{\GN\op}(L^2(\GN),\H)\rlh\GN, $$
equipped with the obvious left $\GM$ action $A(R)=AR$, right $\GN$
action $(R)B=RB$, and  $\GN$-valued inner product $
\la A,B\ra_{\GN}=A^*B$, where $\GN$ is identified with its (left) \rep\ on $L^2(\GN)$, is a
normal selfdual $\GM$-$\GN$ Hilbert bimodule.
\item Conversely, let $\GM\rac\CE\rlh\GN$ be a normal
 selfdual $\GM$-$\GN$ Hilbert bimodule. Then
$\GM\rac \CE\hat{\ot}_{\GN}L^2(\GN)\lac\GN$,
equipped with the obvious left $\GM$ action, and the right $\GN$
action inherited from its canonical right action on $L^2(\GN)$
(cf.\ Remark \ref{TT}), is an $\GM$-$\GN$ correspondence.
Here the Hilbert space $\CE\hat{\ot}_{\GN}L^2(\GN)$ is the 
interior
tensor product of $\CE\rlh\GN$ and $\GN\rac L^2(\GN)\rlh\C$,
the right $\GN$ action on the latter passing to the quotient in the
obvious way.
\item 
Up to isomorphism, the above passage from correspondences
to normal selfdual Hilbert bimodules (and back) maps $\boxtimes_{\GN}$ into
$\hat{\ot}_{\GN}$ (and back).
\item 
The maps in items 1 and 2 establish an isomorphism between \Wa\ and the subcategory
of \Ca\ consisting of \vna s as objects and normal selfdual Hilbert bimodules as arrows.
\end{enumerate}
\end{proposition}

Before giving the proof, let us note that the passage from $\H$ 
in claim 1 to $\Hom_{\GN\op}(L^2(\GN),\H)$  is not a big deal, since by
Lemma
\ref{saulem} the latter may be identified with the dense subspace
$\til{\H}\subset\H$. However, if one formulates the proposition in
terms of $\til{\H}$, with inner product
$\la\Ps,\Ph\ra_{\GN}=R^*_{\Ps}R_{\Ph}$, one should be aware that the
right $\GN$ action $\pi_R$ on $\til{\H}$ should be
$\pi_R(B)\Ps=\Ps\sg_{i/2}(B)$, in terms of the given right $\B$ action
on $\H$. Here $\sg_t(\cdot)$ is the modular group acting on $\GN$ as
defined in the Tomita-Takesaki theory. The correction factor
$\sg_{i/2}$ is needed to satisfy the compatibility condition
(\ref{comp}).  For finite \vna s this correction may be ignored, as
the modular group is trivial.
\begin{Proof}
The first construction in Proposition \ref{BDH2.2} is a special case
of Thm.\ 6.5 in \cite{Rie2}, which guarantees selfduality and
normality; the fact that the inner product is well defined is already
clear from Lemma \ref{saulem}. 

To show that the second map defined in Proposition \ref{BDH2.2} is
an inverse of the first up to equivalence, we find a unitary from
$\Hom_{\GN\op}(L^2(\GN),\H)\hat{\ot}_{\GN}L^2(\GN)$ to $\H$ that
intertwines the given $\GM$ and $\GN$ actions.  We restrict the proof to
the $\sg$-unital case. Since $\CE$ is dense
in $\CE\hat{\ot}_{\GN}L^2(\GN)$ under the identification
$\Ps\mapsto\Ps\ot_{\CN}\Om$ for any $\CE$, for
$\CE=\Hom_{\GN\op}(L^2(\GN),\H)$ we may initially define the map
in question by $R\ot_{\GN}\Om\mapsto R\Om$. 
In the picture of $\Hom_{\GN\op}(L^2(\GN),\H)$ as $\til{\H}$, and
of $\til{\H}$ as a (dense) subspace of $\til{\H}\hat{\ot}_{\GN}L^2(\GN)$
under the identification $\Ps\mapsto\Ps\ot_{\GN}\Om$, our map is
just the identity, since $R_{\Ps}\Om=\Ps$.

Since
$$
\| R\ot_{\GN}\Om\|^2= (\Om,\la R,R\ra_{\GN}\Om)_{L^2(\GN)}=
(\Om,R^*R\Om)_{L^2(\GN)}= (R\Om,R\Om)_{\H}=
\| R\Om\|^2_{\H},$$
our map is isometric, hence injective. Its image is dense by Lemma
\ref{saulem}, so that we obtain a unitary operator after extension
by continuity. It is trivial that the map intertwines the left $\GM$ actions.
For the right $\GN$ actions, take $B\in\GN$ and note that $$
(R\ot_{\GN}\Om)B= R\ot_{\GN}(\Om B)\mapsto R(\Om B)=(R\Om)B,$$
since $R\in \Hom_{\GN\op}(L^2(\GN),\H)$. Here $\Om B=JB^*\Om$.

To show that the first map defined in Proposition \ref{BDH2.2} is
an inverse of the second up to equivalence, we find a map
$T:\CE\raw\Hom_{\GN\op}(L^2(\GN),\CE\hat{\ot}_{\GN}L^2(\GN))$
that is adjointable, isometric with respect to the $\GN$-valued
inner products in question, and intertwines the given $\GM$ and $\GN$ actions.
We first note that for $\Ps\in\CE$ the vector
$\Ps\ot_{\GN}\Om$ lies in $\widetilde{\CE\hat{\ot}_{\GN}L^2(\GN)}$
(cf.\ Lemma \ref{saulem}), since  for $A\in\GN$ one has
\begin{eqnarray*}
\| R_{\Ps\ot_{\GN}\Om}(\Om A)\|^2 & = & 
\|(\Ps\ot_{\GN}\Om)A\|_{\CE\hat{\ot}_{\GN}L^2(\GN)}^2=
\|\Ps\ot_{\GN}(\Om A)\|_{\CE\hat{\ot}_{\GN}L^2(\GN)}^2 \\
& = &
(\Om A, \la\Ps,\Ps\ra_{\CN}^{\CE}\Om A)_{L^2(\GN)}\leq
\|\la\Ps,\Ps\ra_{\CN}^{\CE}\| \; \|\Om A\|^2.
\end{eqnarray*}

Thus we may define $T$ by $T:\Ps\mapsto R_{\Ps\ot_{\GN}\Om}$.
If we identify $\CE$ with a dense subspace of
$\CE\hat{\ot}_{\GN}L^2(\GN)$ by $\Ps\mapsto\Ps\ot_{\GN}\Om$,
and identify $\Hom_{\GN\op}(L^2(\GN),\CE\hat{\ot}_{\GN}L^2(\GN))$
with the dense subspace $\widetilde{\CE\hat{\ot}_{\GN}L^2(\GN))}$
of $\CE\hat{\ot}_{\GN}L^2(\GN))$ as in Lemma \ref{saulem}, which
dense subspace is precisely $\CE$, we see that $T$ does just nothing.
In any case,  note that
\begeq
\la R_{\Ps\ot_{\GN}\Om},R_{\Ph\ot_{\GN}\Om}\ra^{\Hom(\ldots)}_{\GN}=
\la\Ps,\Ph\ra_{\GN}^{\CE}. \label{beq}
\end{equation}
This may be verified by identifying $\GN$ with $\GN\rac L^2(\GN)$, and
taking matrix elements between vectors in the dense set $\Om\GN$: 
the left-hand side of (\ref{beq}) is $R_{\Ps\ot_{\GN}\Om}^*R_{\Ph\ot_{\GN}\Om}$
by definition, and
one has
\begin{eqnarray*}
(\Om B, R_{\Ps\ot_{\GN}\Om}^*R_{\Ph\ot_{\GN}\Om}\Om A)_{L^2(\GN)}& = &
((\Ps\ot_{\GN}\Om)B,(\Ph\ot_{\GN}\Om)A)_{\CE\hat{\ot}_{\GN}L^2(\GN)}\\
 = 
(\Ps\ot_{\GN}(\Om B),\Ph\ot_{\GN}(\Om A))_{\CE\hat{\ot}_{\GN}L^2(\GN)}& = &
(\Om B, \la\Ps,\Ph\ra_{\GN}^{\CE}\Om A)_{L^2(\GN)}.
\end{eqnarray*}

Since (\ref{beq}) may be read as $$
\la R_{\Ps\ot_{\GN}\Om},T\Ph\ra^{\Hom(\ldots)}_{\GN}=
\la T^* R_{\Ps\ot_{\GN}\Om},\Ph\ra_{\GN}^{\CE},
$$ with $T^*  R_{\Ps\ot_{\GN}\Om}=\Ps$,
we see that $T$ is adjointable.
It is trivial to verify that it has the correct intertwining properties
as well.

Finally, we prove claim 3, from which no.\ 4 is obvious. Given normal selfdual Hilbert bimodules
$\GM\rac\CE\rlh\GN$ and $\GN\rac\CF\rlh\GP$, we put
\bea
\H_1 & = & (\CE\hat{\ot}_{\GN}\CF)\hat{\ot}_{\GP}L^2(\GP);\\
\H_2 & = & (\CE\hat{\ot}_{\GN}L^2(\GN))\boxtimes_{\GN} 
(\CF\hat{\ot}_{\GP}L^2(\GP)),
\eea
and show that the correspondences $
\GM\rac \H_1\lac\GP$ and $\GM\rac \H_2\lac\GP$ are isomorphic.
To do so, define a map from $\H_1$ to $\H_2$ by
$$
(\Ps\ot_{\GN}\Ph)\ot_{\GP}\Om_{\GP}\mapsto 
(\Ps\ot_{\GN}\Om_{\GN})\boxtimes_{\GN}(\Ph\ot_{\GP}\Om_{\GP}).
$$
A simple computation shows that this map is isometric, and extends to
a surjective, hence unitary operator, which intertwines the given $\GM$ and
$\GN$ actions. \end{Proof}

Proposition \ref{BDH2.2} leads to a third description of Connes's
tensor product, which is a mixture of the previous two descriptions. The
tensor product $$\GM\rac
\H\boxtimes_{\GN}\CK\lac\GP$$ of the correspondences
$\GM\rac\H\lac\GN$ and $\GN\rac\CK\lac\GP$ is the interior  tensor
product $$\GM\rac \Hom_{\GN\op}(L^2(\GN),\H)\hat{\ot}_{\GN}\CK\rlh\C$$
of the Hilbert bimodules $\GM\rac \Hom_{\GN\op}(L^2(\GN),\H)\rlh\GN$
and $\GN\rac\CK\rlh\C$, where one has to remark separately that the
$\GP$-action on $\CK$ quotients to a well-defined action on the
Hilbert space of the interior tensor product.

Finally, we note two special cases of Proposition
\ref{BDH2.2}.
\begin{example}\label{BDHex}
\begin{enumerate}
\item A \Hs\ $\H$ is both an $\GM$-$\C$
correspondence and a normal selfdual $\GM$-$\C$ Hilbert bimodule, where
$\GM\subseteq\BH$. The maps defined in Proposition
\ref{BDH2.2}.1 and 2, then, act trivially on \Hs s. 
\item
The correspondence $\GM\rac L^2(\GM)\lac\GM$ is mapped into the
Hilbert bimodule $\GM\rac\GM\rlh\GM$ of Example \ref{BB}, and vice
versa. This is because $$
\Hom_{\GM\op}(L^2(\GM),L^2(\GM))=\GM''=\GM.$$
\end{enumerate}
\end{example}

\section{Groupoids}
Recall the notation for categories explained at the end of the
Introduction.  A groupoid is a small category in which each arrow is
invertible (i.e., an isomorphism). For categories that are groupoids
we use the generic symbol $G$ rather than $C$ (as well as $H$, $K$).
We denote the inverse by $I:G_1\raw G_1$, and the source and target
maps by $s,t:G_1\raw G_0$. The object space $G_0$ is usually regarded
as a subspace of $G_1$.

Three examples of groupoids that should always be kept in mind are groups
$G$ (where $G_1=G$ and $G_0=\{e\}$), sets $S$ (where $G_1=G_0=S$ with the
obvious trivial groupoid structure),
and pair groupoids over a set $S$; here one has $G_1=S\x S$ and
$G_0=S$, with  $s(x,y)=y$, $t(x,y)=x$, $(x,y)\inv=(y,x)$, 
$(x,y)(y,z)=(x,z)$, and $1_x=(x,x)$.

We will often use the notation $$
A\times^{f,g}_B C= \{(a,c)\in A\x C\mid f(a)=g(c)\}
$$
for the fiber product
of sets $A$ and $C$ with respect to maps $f:A\raw B$ and $g:C\raw B$.
When the maps $f$ and $g$ are obvious, we simply write $A\times_B C$.
\subsection{The category \Gr\ of groupoids and principal bibundles}\label{MM1}
Purely algebraic groupoids may be organized into a category by
regarding them as categories themselves. The most obvious arrows
between categories are functors, but the category whose objects are
categories and whose arrows are functors entails a notion of
isomorphism that category theorists frown upon. The classy notion
of isomorphism between categories is that of (natural) equivalence,
which is achieved by taking the arrows between categories to be
equivalence classes of functors under natural isomorphism.

A (seemingly) different category of groupoids is obtained by taking
the arrows to be so-called principal bibundles.  Historically,
principal bibundles, seen as generalized maps between topological or
Lie groupoids, were introduced by Skandalis for holonomy groupoids of
foliations; see \cite{Hae,HS}.  Independently, Moerdijk \cite{Moe}
defined such bibundles in the context of
topos theory.  The connections between functors, principal bibundles,
and associated notions of Morita equivalence were elaborated by
Moerdijk and Mr\v{c}un \cite{Moe,Mrc1,Mrc2,vdlaan}, on whose work 
sections \ref{MM1}, \ref{MM2}, \ref{MM3}, and \ref{MM4} are largely based.
\begin{definition}\label{defGr}
The category \Grb\ has groupoids as objects, and  isomorphism
classes of functors as arrows. Composition is defined by $[\Ps]\circ
[\Ph]=[\Ps\circ \Ph]$, and the unit arrow at a groupoid $G$ is
$1_G=[\id_G]$, where $\id_G:G\raw G$ is the identity functor.
\end{definition}

For purely algebraic groupoids, there is an isomorphic way of defining
\Grb\ through principal bibundles. Since this formulation will play a
central role in the discussion of Lie groupoids, we review the
necessary definitions of actions and bibundles for groupoids (the
former notion goes back to Ehresmann, cf.\ \cite{Mac}).
\begin{definition}\label{Gaction} 
\begin{enumerate}
\item
Let $G$ be a  groupoid and $M$  a set equipped with
a ``base map'' $M\stackrel{\ta}{\raw} G_0$.
  A left $G$ action on $M$ (more precisely, on $\ta$) is a 
map $(x,m)\mapsto xm$ from $G_1 \times^{s,\ta}_{G_0}M$ to $M$,
 such
that $\ta(xm)=t(x)$, $xm=m$ for all $x\in G_0$, and $x(ym)=(xy)m$
whenever $s(y)=\ta(m)$ and $t(y)=s(x)$.  
\item
Given a base map $M\stackrel{\rh}{\raw} H_0$, a right action of a
groupoid $H$ on $M$ (or $\rh$) is a map $(m,h)\mapsto mh$ from
$M\times^{\rh,t}_{H_0} H_1$ to $M$ that satisfies $\rh(mh)=s(h)$, $mh=m$
for all $h\in H_0$, and $(mh)k=m(hk)$ whenever $\rh(m)=t(h)$ and
$t(k)=s(h)$.
\item
A $G$-$H$ bibundle $M$  carries a left
$G$ action as well as a right $H$-action that commute. That is,
 one has
$\ta(mh)=\ta(m)$, $\rh(xm)=\rh(m)$, and $(xm)h=x(mh)$ 
whenever defined.  On occasion, we simply write $G\rac M\lac H$.
\end{enumerate}
\end{definition}

We will be interested in special bibundles, called principal \cite{Hae,HS,Moe,Mrc1,Mrc2}.
\begin{definition}\label{defprin}
A (left)  $G$ bundle $M$ over a set $X$ consists of
a (left) $G$ action on $M$ and a  map $\pi:M\raw X$ that
is invariant under the $G$ action. Similarly for right actions.

A (left) $G$ bundle $M$ over $X$ is called principal when $\pi$ is 
surjective, and the $G$
action is free (in that $xm=m$ iff $x\in G_0$) and transitive along
the fibers of $\pi$.

 A $G$-$H$ bibundle $M$ is called left principal when it is principal for the
 $G$ action with respect to $X=H_0$ and $\pi=\rh$.
Similarly, it is called right principal when it is principal for the
 $H$ action with respect to $X=G_0$ and $\pi=\ta$, and biprincipal
when it is principal for both the left and the right action.
\end{definition}

Note that this definition of a principal action is different from the
one in \cite{MRW}.  

Two  $G$-$H$ bibundles $M,N$ are called isomorphic
if there is a bijection $M\raw N$
that intertwines the maps $M\raw G_0$, $M\raw H_0$ with the maps
$N\raw G_0$, $N\raw H_0$, and in addition intertwines the $G$ and $H$
actions (the latter condition is well defined because of the former).
\begin{proposition}\label{smbi}
There is a  bijective correspondence between isomorphism
classes of functors $\Ph:G\raw H$ 
and isomorphism classes of right principal $G$-$H$ bibundles.
\end{proposition}
\begin{Proof}
 Given $\Ph:G\raw H$, we define a $G$-$H$ bibundle $M_{\Ph}$ by
 putting $$M_{\Ph}=G_0\times^{\Ph_0,t}_{H_0}H_1,$$ with base maps
 $\ta:M_{\Ph}\raw G_0$ given by $\ta(u,h)=u$ and $\rh:M_{\Ph}\raw H_0$
 defined by $\rh(u,h)=s(h)$.  The left $G$ action is
 $x(u,h)=(t(x),\Ph_1(x)h)$, and the right $H$ action is
 $(u,h)k=(u,hk)$.

It is clear that $M_{\Ph}$ is right principal. If $\Ps:G\raw H$ is
naturally isomorphic to $\Ph$ through $\nu:G_0\raw H_1$ (in that
$\Ph_0(u)=\nu_u(\Ps_0(u))$, natural in $u$), then the map
$(u,h)\mapsto (u,\nu_u\inv h)$ from $M_{\Ps}$ to $M_{\Ph}$ 
is an isomorphism of bibundles.

Conversely, given a right principal $G$-$H$ bibundle $M$, we use the
axiom of choice to pick a section $\sg:G_0\raw M$ of $\ta$, and define
a functor $\Ph^{\sg}:G\raw H$ by $\Ph^{\sg}_0(u)=\rh\circ\sg(u)$,
and defining  $\Ph^{\sg}_1(x)$ as the unique element of $H_1$ 
that satisfies $x\sg(s(x))=\sg(t(x))\Ph^{\sg}_1(x)$.

A different section $\til{\sg}$ of $\ta$ is related to $\sg$ through
$\til{\sg}(u)=\sg(u)\nu\inv_u$; the map $\nu:G_0\raw H_1$ is then
a natural isomorphism from $\Ph^{\sg}$ to $\Ph^{\til{\sg}}$.

Applying this procedure to $M=M_{\Ph}$ as defined above, and choosing
the section $\sg(u)=(u,\Ph_0(u))$, it follows that $\Ph^{\sg}=\Ph$.
Hence by the previous paragraph an arbitrary section will lead to a
functor naturally isomorphic to $\Ph$. 

We have $M\simeq M_{\Ph^{\sg}}$ as $G$-$H$ bibundles through
the map $m\mapsto (\ta(m),h)$, where $h$ satisfies $m=\sg(\ta(m))h$
(and is uniquely defined by this property by right principality).

Finally, we show that isomorphic right principal bibundles induce
isomorphic functors. For given functors $\Ph,\Ps:G\raw H$, suppose
that $M_{\Ph}\simeq M_{\Ps}$ as $G$-$H$ bibundles. Such an isomorphism
$M_{\Ph}\raw M_{\Ps}$ is necessarily of the form $(u,h)\mapsto (u,\nu_u h)$
for some $\nu:G_0\raw H_1$ that defines an isomorphism from $\Ph$ to $\Ps$
(the naturality of the latter isomorphism follows from the requirement
of $G$ equivariance of the former isomorphism). If $M\simeq M'$, then,
as we have seen, $M\simeq M_{\Ph^{\sg}}$ and $M'\simeq M_{\Ph^{\sg'}}$ 
for any choice of sections, hence $M_{\Ph^{\sg}}\simeq M_{\Ph^{\sg'}}$,
so that $\Ph^{\sg}\simeq \Ph^{\sg'}$ by the previous result.
\end{Proof}

Suppose one has right principal bibundles $G\rac M\lac H$ and $H\rac N\lac K$.
The fiber product $M\times_{H_0} N$ 
carries a right $H$ action, given by
$h:(m,n)\mapsto (mh,h\inv n)$ (defined as appropriate). We denote the
 the orbit space by
\begeq
M\otg_H N=(M\times_H N)/H. \label{btp} 
\end{equation}
This is a $G$-$K$ bibundle under the obvious maps $\til{\ta}:M\otg_H N\raw
G_0$ and $\til{\rh}:M\otg_H N\raw K_0$, viz.\ 
$\til{\ta}([m,n]_H)=\ta(m)$ and $\til{\rh}([m,n]_H)=\rh(n)$, left $G$ action
given by $x[m,n]_H=[xm,n]_H$, and right $K$ action defined by 
$[m,n]_Hk=[m,nk]_H$. 

\begin{lemma}\label{GGlem}
Define the canonical $G$-$G$ bibundle $G$ by putting
$M=H=G$, $\ta=t$, and $\sg=s$ in the above definitions; the left and right
actions are simply given by multiplication in the groupoid. 
This
 bibundle is a left and a right unit for the
bibundle tensor product (\ref{btp}), up to isomorphism.
\end{lemma}

\begin{Proof} 
For any $G$-$H$ bibundle $M$ the map $G\otg M\raw M$ given by
$[x,m]_G\mapsto xm$ is an isomorphism, etc.
\end{Proof}

\begin{corollary}\label{GrGrb}
The category \Grb\ is isomorphic to the category \Gr\ having groupoids
as objects and isomorphism classes of right principal bibundles as
arrows, composed by (\ref{btp}), descending to isomorphism
classes. The units in \Gr\ are the
isomorphism classes of the canonical bibundles.
\end{corollary}
\begin{Proof}
Taking $H=G$ and $\Ph=\id$ in Proposition \ref{smbi},  one finds that $M_{\Ph}$ is
isomorphic to the canonical $G$-$G$ bibundle $G$.
If $\Ph:G\raw H$ and $\Ps:H\raw K$ are functors,
 simple computation yields
\begin{equation}
M_{\Ph}\otg_H M_{\Ps}\simeq M_{\Ps\circ\Ph},
\end{equation}
so that $[M_{\Ps}]\circ [M_{\Ph}]=[M_{\Ps\circ\Ph}]$ in \Gr.
\end{Proof}

This result may, indeed, serve as the motivation for the construction
(\ref{btp}).
\subsection{Morita equivalence for  groupoids}\label{MM2}
The standard definition of Morita equivalence for groupoids is as follows
\cite{Hae,MRW,X1}.
\begin{definition}\label{MEGR}
A right principal $G$-$H$ bibundle $M$  is called
an equivalence bibundle when it is biprincipal.
Two groupoids related by an equivalence bibundle are called Morita
equivalent.
\end{definition}

The groupoid analogue of Proposition \ref{algmor} is
\begin{proposition}\label{gromor}
The following conditions are equivalent:
\begin{enumerate}
\item
A $G$-$H$ bibundle $M$ is an equivalence bibundle;
\item 
The isomorphism
class $[M]\in (G,H)$ is invertible as an arrow in \Gr;
\item The isomorphism class $[\Ph]\in\Grb$ that corresponds to $[M]\in\Gr$
(see Proposition \ref{smbi}) is invertible as an arrow in \Grb;
\item
Each  functor in the above class $[\Ph]\in\Grb$  defines a category equivalence.
\end{enumerate}

Hence two groupoids are Morita equivalent iff they are isomorphic objects
in \Grb\ or in \Gr\ and iff they are equivalent as categories.
\end{proposition}

Recall \cite{MacLane} that a functor $\Ph:G\raw H$ is a category equivalence
when $\Ph_0$ is essentially surjective (i.e., for each $v\in H_0$
there is an $u\in G_0$ for which $\Ph_0(u)\cong v$) and $\Ph_1$
is full (in that for all $u,u'\in G_0$ the map
$\Ph_1: (u,u')\raw (\Ph_0(u),\Ph_0(u'))$ is surjective) as well as
faithful (i.e., the above map is injective). One then says that
$\Ph$ is essentially surjective on objects and fully faithful on arrows.
\begin{Proof}
It is possible to prove the equivalence of claims 1 and 2 directly;
see the proof of Proposition \ref{MEGR2} below.  
 We just mention that the
inverse of a biprincipal bibundle $G\rac M\lac H$ is simply $H\rac
M\lac G$, with the same base maps, but left and right actions swapped
by composing the original actions with the inverse, for both $G$ and $H$.

Here, we simply refer
to the proof of Proposition \ref{smbi}, adding the following
\begin{lemma}\label{Mrclem}
$M_{\Ph}$ is biprincipal iff $\Ph$ is a category equivalence.  
\end{lemma}
\begin{Proof}
The
surjectivity of $\rh$ corresponds to $\Ph_0$ being essentially
surjective, the freeness of the $G$ action corresponds to the
faithfulness of $\Ph_1$, and the transitivity of the $G$ actions on
the fibers of $\rh$ corresponds to the fullness of $\Ph_1$.  
\end{Proof}

This lemma proves the equivalence of nos.\ 1 and 4 in Proposition
\ref {gromor}. Claims 3 and 4 are equivalent
by a well-known argument in category theory using the axiom of choice
(see \cite{MacLane} or the proof of Proposition \ref{MorMG} below),
and Corollary
\ref{GrGrb} then implies the equivalence of 2 and 3.
\end{Proof}

We now write down the  groupoid counterpart of Definitions
\ref{kejo}, \ref{repsca}, and \ref{repvn}.
\begin{definition}\label{Repgoid}
The \rep\ category $\Rep(G)$ of a groupoid $G$ has
left $G$ actions as objects.  An arrow 
 between an action on
$M\stackrel{\ta}{\raw} G_0$ and one on $N\stackrel{\rh}{\raw} G_0$
is a  map $\phv:M\raw N$ that satisfies $\rh\phv=\ta$
and intertwines the $G$-action.
\end{definition}
As in Proposition \ref{morita}, we now  have
\begin{proposition} \label{Morgr}
If two  groupoids are \Me,
then their \rep\ categories are equivalent.
\end{proposition}
\begin{Proof}
For the category \Gr\ the proof is the same as for algebras, with the
usual modifications.  For \Grb\ one constructs a functor $F$ from
$\Rep(G)$ to $\Rep(H)$ from a category equivalence $\Ps:H\raw G$ as
follows: on objects one maps a left $G$ space $M\stackrel{\ta}{\raw}
G_0$ to a left $H$ space $H_0\times_{G_0}^{\Ps_0,\ta}M$, equipped with the obvious base
map $(u,m)\mapsto u$ and the left $H$ action
$h(u,m)=(t(h),\Ps_1(h)m)$, defined when $u=s(h)$.  On arrows one
extends in the obvious way, i.e., $\phv:M\raw N$ induces $(u,m)\mapsto
(u,\phv(m))$. This functor is a category equivalence, for an inverse
(up to natural isomorphism) is found by picking an inverse $\Ph:G\raw
H$ of $\Ps$ in \Grb\ (i.e., an inverse up to natural isomorphism), and defining a
functor from $\Rep(H)$ to $\Rep(G)$ given on objects by mapping
$M\stackrel{\rh}{\raw} H_0$ to $G_0\times_{H_0}^{\Ph_0,\rh}M$, 
 etc. Indeed, for $M\in\Rep(H)_0$ one finds an
isomorphism $$
H_0\times_{G_0}^{\Ps_0,\id}(G_0\times^{\Ph_0,\rh}_{H_0}M)\simeq M$$ as
left $H$ spaces, given by $(u,v,m)\mapsto \ta_um$, where
$\ta:H_0\raw H_1$ is a natural isomorphism from $\Ph\Ps$ to $\id_H$.
Similarly in the opposite direction. \end{Proof}

The claim fails in the opposite direction, since the right $G$ action
on $G$  cannot be transferred to $F_0(1_G)$; cf.\ the proof of
Proposition \ref{morita}.
\subsection{The category \MG\ of measured groupoids and  functors}
The concept of a measured groupoid emerged from the work of Mackey on
ergodic theory and group \rep s \cite{Mackey}. For the technical
development of this concept see \cite{Ram1,Hah1,FHM}. A different
approach was initiated by Connes \cite{Conmg}. The connection between
measured groupoids and locally compact groupoids is laid out in
\cite{Ren,Ram2}. 
\begin{definition}\label{defmg}
A Borel groupoid is a groupoid $G$ whose total space $G_1$ is an
analytic Borel space, such that $I:G_1\raw G_1$ is a Borel map,
$G_2\subset G_1\x G_1$ is a Borel subset, and multiplication
$m:G_2\raw G_1$ is a Borel map.  It follows that $G_0$ is a Borel set
in $G_1$, and that $s$ and $t$ are Borel maps.

A left Haar system on a Borel groupoid is a family of measures
$\{\nu^u\}_{u\in G_0}$, where $\nu^u$ is supported on 
the $t$-fiber $G^u=t\inv(u)$, which is left-invariant in that
\begin{equation}
\int d\nu^{s(x)}(y)\, f(xy)=\int d\nu^{t(x)}(y)\, f(y) \label{mg2}
\end{equation}
for all $x\in G_1$ and all positive 
Borel functions $f$ on $G_1$ for which both
sides are finite.

A measured groupoid is a Borel groupoid equipped with a Haar system
as well as a Borel measure $\til{\nu}$ on $G_0$ with the property that
the measure class of the measure $\nu$ on $G_1$, defined by
\begin{equation}
\nu=\int_{G_0} d\til{\nu}(u)\, \nu^u, \label{mg1}
\end{equation}
is invariant under $I$ (in other words, $I(\nu)\sim \nu$).
\end{definition}

Recall that the push-forward of a measure under a Borel map
is given by $t(\nu)(E)=\nu(t\inv(E))$ for Borel sets $E\subset G_0$.

This definition turns out to be best suited for categorical
considerations. It differs from the one in \cite{Ram1,Hah1}, which is
stated in terms of measure classes. However, the measure class of
$\nu$ defines a measured groupoid in the sense of \cite{Ram1,Hah1},
and, conversely, the latter is also a measured groupoid according to
Definition \ref{defmg} provided one removes a suitable null set from
$G_0$, as well as the corresponding arrows in $G_1$; cf.\ Thm.\ 3.7 in
\cite{Hah1}. Similarly, Definition \ref{defmg} leads to a locally
compact groupoid with Haar system \cite{Ren} after removal of such a
set; see Thm.\ 4.1 in \cite{Ram2}.  A measured groupoid according to
Connes \cite{Conmg} satisfies Definition \ref{defmg} as well, with
$\til{\nu}$ constructed from the Haar system and a transverse measure
\cite{MoS}. See all these references for extensive information and examples.

The fact that a specific choice of a measure in its class is made in
Definition \ref{defmg} is  balanced by the concept of a
measured functor between measured groupoids, which is entirely
concerned with measure classes rather than individual measures.
Moreover, one merely uses the measure class of $\til{\nu}$.

The measure $\til{\nu}$ on $G_0$ induces a measure $\hat{\nu}$ on
$G_0/G$, as the push-forward of $\til{\nu}$ under the canonical
projection.  Similarly for a measured groupoid $H$, for whose measures
we will use the symbol $\lm$ instead of $\nu$.  A functor $\Ph:G\raw
H$ that is a Borel map induces induces a Borel map $\hat{\Ph}_0:G_0/G\raw H_0/H$.
\begin{definition}\label{defmf}
A measured functor $\Ph:G\raw H$ between two measured groupoids is
a Borel map that is algebraically a functor and satisfies
$\hat{\Ph}_0(\hat{\nu})\prec\prec \hat{\lm}$.
\end{definition}

The latter condition means that $\hat{\lm}(E)=0$ implies
$\hat{\nu}(\hat{\Ph}_0\inv(E))=0$ for all Borel sets $E\subset
H_0/H$, or, equivalently, that $\til{\lm}(F)=0$ implies
$\til{\nu}(\Ph_0\inv(F))=0$ for all saturated Borel sets $F\subset H_0$
(saturated means that if a point lies in $F$ then all points isomorphic to
it must lie in $F$ also). This requirement excludes a number of
pathologies, but includes certain desirable functors that would be
thrown out if the more restrictive condition $\Ph_0(\til{\nu})\prec\prec
\til{\lm}$ had been used. For example, any inclusion of the trivial
groupoid (consisting of a point) into, say, the pair groupoid over
$\R$ with Lebesgue measure, is now a measured functor.

What we here call a measured functor is called a strict homomorphism
in \cite{Ram1}, and a homomorphism in \cite{Ram2}.  Also, note that in
\cite{Mackey,Ram1,FHM} various more liberal definitions are used (in
that one does not impose that $\Ph$ be a functor algebraically at all
points), but it is shown in \cite{Ram2} that if one passes to natural
isomorphism classes, this greater liberty gains little.
\begin{definition}\label{defMG}
The category \MG\ has measured groupoids as objects, and  isomorphism
classes of measured functors as arrows. 
(Here a natural transformation $\nu:G_0\raw H_1$ between Borel functors
from $G$ to $H$ is required to be a Borel map.) Composition and units
are as in Definition \ref{defGr}.
\end{definition}

This definition is a direct adaptation of the category \Grb\ defined in
the purely algebraic case. It is possible to define a counterpart of
\Gr\ for measured groupoids as well, but this does not appear to be
very useful.  In the smooth case, it will be the other way round.
\subsection{Morita equivalence for measured groupoids}
The definition of Morita equivalence for measured groupoids will
be adapted from the notion of an equivalence of categories.
\begin{definition}\label{defMEMG}
A measured functor $\Ph:G\raw H$ between measured groupoids is
called a measured equivalence functor when $\Ph$ is algebraically
an equivalence of categories (i.e., $\Ph_0$ is essentially surjective
and $\Ph_1$ is fully faithful) and 
$\hat{\Ph}_0([\hat{\nu}])=[\hat{\lm}]$.
Two measured groupoids are called Morita equivalent when they are related
by a measured equivalence functor.
\end{definition}

As before, these concepts turn out to be the same as invertibility
and isomorphism in the pertinent category.
\begin{proposition}\label{MorMG}
A measured functor $\Ph:G\raw H$ is a measured equivalence functor iff
its isomorphism class $[\Ph]\in(G,H)$ is invertible as an arrow
in \MG.

In other words, two measured groupoids are Morita equivalent iff they
are isomorphic objects in \MG.
\end{proposition}

This proposition shows that our definition of Morita equivalence is
the same as the notion of strict similarity in \cite{Ram1}, and 
somewhat clarifies this notion. 
\begin{Proof}
The proof hinges on the measurable version of the axiom of choice for
analytic sets (cf., e.g., \cite{KR2}, Thm.\ 14.3.6), which we recall without proof.
\begin{lemma}\label{MAC}
Let $X$ and $Y$ be Polish spaces with associated Borel structure, and
let $Z\subset X\x Y$ be analytic. Let $$\til{Y}=\{y\in Y\mid
\exists x\in X, (x,y)\in Z\}.$$
Then there exists a Borel map $g:\til{Y}\raw X$ such that $(g(y),y)\in
Z$ for all $y\in\til{Y}$.
\end{lemma}

Now suppose  $\Ph: G\raw H$ is an equivalence functor.
In the lemma, take $X=G_0$, $Y=H_0$, and $$Z=\{(u,v)\in G_0\x H_0\mid
\Ph_0(u)\cong v\}.$$
Note that $Z=\coprod_{\CO} \Ph_0\inv(\CO)\x\CO$, where the disjoint
union ranges over all $H$ orbits $\CO$ in $H_0$.  Now
$\CO=t(s\inv(v))$ for any $v\in\CO$; in a Polish space points are
Borel sets, hence $s\inv(v)$ is Borel, so that $\CO$ is analytic. As
the disjoint union of analytic sets, $Z$ is analytic as well.  Note
that $\til{Y}=H_0$, as $\Ph_0$ is essentially surjective.  Choosing
some $g$ as in the lemma, we may define $\Ps_0=g:H_0\raw G_0$.

To define $\Ps_1:H_1\raw G_1$, take $X=H_1$, $Y=H_0$, and $$
Z=\{(x,v)\in H_1\x H_0\mid x\in (\Ph_0\Ps_0(v),v)\}.$$ in Lemma
\ref{MAC}. Then $$ Z=\coprod_{v\in H_0} \{s\inv(v)\cap
t\inv(\Ph_0\Ps_0(v)),v\}, $$ which is a Borel set, hence
analytic. Using Lemma \ref{MAC} once again, it follows that there
exists a Borel map $g:H_0\raw H_1$, in terms of which $\Ps_1$ can be
defined as in the purely algebraic case
\cite{MacLane}: since $\Ph_1$ is fully faithful, for given
$x\in (v',v)\subset H_1$  there is a unique $h\in (\Ph_0\Ps_0(v'),
\Ph_0\Ps_0(v)) \subset G_1$ for which $g(v')xg(v)\inv=\Ph_1(h)$.
One then puts $\Ps_1(x)=h$; the map $\Ps_1$ thus defined is Borel,
since $g$ and $\Ph_1$ are. As in the purely algebraic case, it follows
that $\Ps\circ\Ph\simeq \id_G$ and $\Ph\circ\Ps\simeq \id_H$
via natural transformations that in the measured case can be chosen to be 
Borel maps. 

It remains to be shown that $\hat{\Ps}_0(\hat{\lm})\prec\prec
\hat{\nu}$. We will, in fact, prove that
\begin{equation}
\hat{\Ps}_0([\hat{\lm}])=[\hat{\nu}]. \label{hulp}
\end{equation}
Denote the saturation of a set $B$ in the base space of some groupoid
by $\CS(B)$; hence $\CS(B)$ consists of all points that are isomorphic
to some point in $B$. It is easy to see that
$\Ph_0\inv(\CS(B))=\CS(\Ps_0(B))$ for any $B\subset H_0$. Similarly,
$\Ps_0\inv(\CS(E))=\CS(\Ph_0(E))$ for all $E\subset G_0$. We know from
the definition of $\Ph$ as an equivalence functor that
$\til{\nu}(\Ph_0\inv(B))=0$ for saturated $B$ is equivalent to
$\til{\lm}(B)=0$. Since $E=\Ph_0\inv(B)$ (which is automatically saturated)
for $B=\CS(\Ph_0(E))$, it follows that 
$\til{\lm}(\Ps_0\inv(E))=0$ for saturated $E$ is equivalent to
$\til{\nu}(E)=0$. This implies (\ref{hulp}).

It follows that $[\Ph]$ is invertible in \MG. The converse implication
is easier, and is left to the reader. \end{Proof}

We leave the formulation of the appropriate measured versions of
Definition \ref{Repgoid} and Proposition \ref{Morgr} to the reader; 
there is no clear need for such results. Indeed, it has either been
the measured groupoids themselves, or the \vna s defined by them \cite{Hah2,CT}
that have been the main objects of study in the literature.
For our purposes, the dichotomy between measured and Lie groupoids is interesting:
the category \MG\ of measured groupoids has been modeled on \Grb, whereas the
category \LG\ of Lie groupoids will be shaped after \Gr. 
\subsection{The category \LG\ of Lie groupoids and principal bibundles}\label{MM3}
 A Lie groupoid is a groupoid for which $G_1$
and $G_0$ are manifolds, $s$ and $t$ are surjective submersions, and
$m$ and $I$ are smooth.  It follows that object inclusion is an immersion, that
$I$ is a diffeomorphism, that $G_2$ is a closed submanifold of $G_1\x
G_1$, and that for each $q\in G_0$ the fibers $s\inv(q)$ and
$t\inv(q)$ are submanifolds of $G_1$.  References on Lie groupoids
that are relevant to the themes in this paper include
\cite{Mac,CW,NPL3}.

Definition \ref{defGr} may be adapted to the smooth setting in the obvious way,
requiring functors and natural transformations to be smooth.
This yields a category $\LGb$ whose objects are Lie groupoids and whose arrows
are isomorphism classes of smooth functors.

We now prepare for the definition of the smooth analogue of \Gr.
In Definition \ref{Gaction} of a groupoid action one now requires
$M$ to be a manifold, and the base maps as well as the maps defining
the action to be smooth. In Definition \ref{defprin} 
 both $M$ and $X$ should be manifolds, and in a principal
bundle $\pi$ has to be a smooth surjective submersion. An equivalent
way of defining a smooth right principal bundle is to require that the map
from $M \times_{H_0}^{\rh,t}H_1\raw M\times_X
M$ given by $(m,h)\mapsto (mh,m)$ be a diffeomorphism.  
The bijection occurring in the definition of isomorphism of bibundles must be
a diffeomorphism in the smooth case.
\begin{lemma}\label{LGL1}
A right principal $G$ action is proper (in that the map $(m,h)\mapsto
(mh,m)$ from $M \times_{H_0}^{\rh,t}H_1\raw M\times M$ is proper), and
$M/H\simeq X$ through $\pi$.  Similarly for a left principal
action.
\end{lemma}
\begin{Proof}
Take an open set $U\subset X$ on which $\pi$ has a smooth
cross-section $\sg:X\raw M$, and note that $\pi\inv(U)\simeq U
\times_{H_0}^{\rh\sg,t} H_1$ in a $H$ equivariant way through
$m\mapsto (\pi(m),h)$, where $h\in H_1$ is uniquely defined by the
property $\sg(\pi(m))h=m$; cf.\ the proof of Proposition \ref{smbi}.
Hence the $H$ action on the right-hand side is $(u,h)k=
(u,hk)$. This implies that the $H$ action on $M$ is proper, since the
$H$ action on itself is. Moreover, one clearly has
$\pi\inv(U)/H\simeq U$. 
\end{Proof}

The proof of the following lemma  is an easy exercise.
\begin{lemma}\label{newlemma}
Let two bibundles $G\rac M\lac H$ and $H\rac N\lac K$
 both be right principal. Then 
their tensor product $M\circledast_H N$ is  a right principal $G$-$K$
bibundle. If the $G$ action on $M$ and the $H$ action on $N$ are proper,
then so is the induced $G$ action on $M\circledast_H N$.

Moreover, the bibundle tensor product (\ref{btp}) between right
principal bibundles is associative up to isomorphism, and passes to
isomorphism classes.
\end{lemma}

Lemma \ref{GGlem} holds for Lie groupoids as well.
Thus one obtains a version of Definition \ref{defCa} for Lie groupoids:
\begin{definition}\label{defLG}
The category \LG\ has Lie groupoids as objects, and isomorphism
classes of right principal bibundles as arrows. The arrows are composed by
the bibundle tensor product (\ref{btp}), for which the canonical bibundles
$G$ are   units.
\end{definition}

If one wishes, one could require the right actions on the bibundles
to be proper; this turns out to be useful for certain purposes.

 Corollary \ref{GrGrb} is not valid for Lie groupoids.  In fact, a
 right principal bibundle $M$ is isomorphic to some $M_{\Ph}$ iff the
 projection $M\raw G_0$ has a smooth section.  Hence the proof of
 Proposition \ref{smbi} breaks down in the smooth case.
The precise connection between \LG\ and \LGb\ will be explained in the
next section.
\subsection{Morita equivalence for Lie groupoids}\label{MM4}
We keep Definition \ref{MEGR} of  also for Lie
groupoids (with the stipulation that the bibundles be smooth, as
explained in the preceding subsection). 
Hence two Lie groupoids are Morita equivalent when they are related
by a biprincipal bibundle.
\begin{proposition}\label{MEGR2}
A right principal bibundle $M$ is an equivalence bibundle
iff its isomorphism class $[M]\in (G,H)$ is invertible as an arrow in
\LG.

In other words, two Lie groupoids are isomorphic objects in \LG\ iff
they are Morita equivalent.
\end{proposition}

It follows from Lemma \ref{newlemma} that this remains true if
the bibundles in the definition of \LG\ are required to be
proper from the right.
\begin{Proof} 
Invertibility of $M$ means that there exists a right principal $H$-$G$
bibundle $M\inv$, such that
\begin{eqnarray}
H\rac M\inv \otg_G M\lac H & \simeq  & H\rac H\lac H;  \label{LM1}\\
G\rac M\otg_H M\inv \lac G & \simeq & G\rac G\lac G. \label{LM2}
\end{eqnarray}

To prove the ``$\Raw$'' claim, take $M\inv$ to be $M$ as a manifold,
seen as a $H$-$G$ bibundle with the same base maps, and left and right
actions interchanged using the inverse in $G$ and $H$. The
isomorphisms (\ref{LM1}) and (\ref{LM2}) are proved by the argument
following Def.\ 2.1 in \cite{MRW}.

For the ``$\Law$'' claim, we first note that  (\ref{LM1}) implies that
$\rh:M\raw H_0$ must be a surjective submersion
 (since the target projection $t:H_1\raw H_0$ is).
Second, (\ref{LM2}) easily implies that 
 the $G$ action on $M$ is free and transitive on the $\rh$-fibers.
 \end{Proof} 

We  now turn to the relationship between the categories
\LGb\ and \LG, which, as we have already  pointed out, are no longer isomorphic 
in the smooth case, or even equivalent. Underlying this difference is
the fact that in the purely algebraic (and also in the measured) case
a functor is a category equivalence (i.e., it is essentially surjective
on objects and fully faithful on arrows) iff it is invertible up to
natural isomorphism \cite{MacLane}, whereas for a smooth functor these
conditions are no longer equivalent.  As in the breakdown of the proof
of Proposition \ref{smbi}, this is because there is no smooth version
of the axiom of choice.  

The notion of isomorphism  of Lie groupoids is, therefore,
coarser in \LG\ than in \LGb. For an example (provided by I. Moerdijk) of two 
Lie groupoids that are  isomorphic in \LG\
but not in \LGb, first note that the pair groupoid over a manifold is
Morita equivalent to the trivial groupoid, both in \LG\ and in \LGb.
Now consider manifolds $P$ and $X$ and a surjective submersion $P\raw
X$.  The restriction of the pair groupoid over $P$ to $P\times_X P$ is
isomorphic to $X$ (seen as a groupoid with $G_0=G_1=X$ having units
only) in \LG: for $M=P$, with obvious actions, is a biprincipal
$(P\times_X P)$-$X$ bibundle.  However, these Lie groupoids are
isomorphic in in \LGb\ iff the fibration $P\raw M$ has a smooth
section.

One can circumvent this problem by a canonical procedure in category theory
\cite{GZ}. Given a category $C$ and a subset $S\subset C_1$ of arrows,
there exists a category $C[S\inv]$ having the same objects as $C$, but
to which formal inverses of elements of $S$ have been added.  There
is, then, a canonical embedding $\iota:C\hookrightarrow
C[S\inv]$.  This new category is characterized by the universal
property that any functor $F:C\raw D$ for which $F_1(x)$ is invertible
in $D$ for all $x\in S$ factors in a unique way as 
\begin{equation}
F=G\circ\iota , \label{GZF}
\end{equation}
 where
$G: C[S\inv]\raw D$ is some functor.  Under certain conditions, summarized
by saying that $S$ allows a calculus of (right) fractions, all arrows in 
$C[S\inv]$ are of the form $\iota(x)\iota(y)\inv$, where $x\in C_1$ and $y\in S$.
\begin{proposition}
Let $S$ be the collection of all smooth category equivalences in \LGb.
The categories $\LGb[S\inv]$ and $\LG$ are isomorphic.
\end{proposition}
\begin{Proof}
In the above paragraph, take $C=\LGb$, $D=\LG$, and $F$ the functor
$F_0=\id$ and $F_1([\Ph]) =[M_{\Ph}]$ appearing in the proof of
Proposition \ref{smbi}. By Proposition \ref{MEGR2} and Lemma
\ref{Mrclem} (which holds also for Lie groupoids),
$F_1(S)$ indeed consists of isomorphisms. 
 For a given  right principal $G$-$H$ bibundle
$M$, let the direct product Lie groupoid $G\x H$ act on $M$ from the
left (with respect to the source map $M\stackrel{\ta\x\rh}{\raw}(G_0\x
H_0)$) by $(x,h)m=xmh\inv$.  Denote the corresponding action groupoid
by $K=(G\x H)\ltimes M$ (see, e.g., \cite{Mac,NPL3,vdlaan}).  Define a functor
$\Upsilon:K\raw H$ by $\Up_0(x,h,m)=\rh(m)$ and $\Up_1(x,h,m)=h$.
Similarly, define a functor $\Om: K\raw G$ by $\Om_0(x,h,m)=\ta(m)$
and $\Om_1(x,h,m)=x$. A straightforward calculation then shows that
$M_{\Om}\otg_G M\simeq M_{\Up}$ as $G$-$K$ bibundles, so that, seen as
arrows in \LG, one has  $[M]\circ [M_{\Om}]=[M_{\Up}]$. Since $\Om$ is
trivially a category equivalence, Proposition \ref{MEGR2} and Lemma
\ref{Mrclem} imply that the arrow
$[M_{\Om}]$ is invertible in \LG, so that
$[M]=[M_{\Up}][M_{\Om}]\inv$.  It follows that
$[M]=G(\iota([\Up])\iota([\Om])\inv)$, where the functor
$G:\LGb[S\inv]\raw\LG$ has been defined in (\ref{GZF}). Hence $G$ is
surjective.

It can be shown that $S$ allows a calculus of right fractions \cite{Moe,vdlaan}. 
With the injectivity of $F:\LGb\raw\LG$, this
 implies that $G$ is injective as well. Hence $G$ is an isomorphism of categories.
\end{Proof}
\begin{corollary}
Two Lie groupoids are Morita equivalent iff they are isomorphic in $\LGb[S\inv]$.
In other words, Morita equivalence of Lie groupoids is
the smallest equivalence relation under which two Lie groupoids related
by a smooth equivalence functor are equivalent.
\end{corollary}

\subsection{The category \SyG\ of symplectic groupoids and symplectic bibundles}
The definition of a suitable category of Poisson manifolds depends
on the theory of symplectic groupoids. These were independently introduced by
Karasev \cite{K0,KM}, Weinstein \cite{W3,CDW,MiWe}, and Zakrzewski
\cite{Zak};
we use the definition of Weinstein (also cf.\ \cite{Vai}).
\begin{definition}\label{defsg}
A symplectic groupoid is a Lie groupoid $\Gm$ for which $\Gm_1$ is a symplectic
manifold, with the property that the graph of $\Gm_2\subset \Gm\x \Gm$
is a Lagrangian submanifold of $\Gm\x \Gm\x \Gm^-$.
\end{definition}

See Lemma \ref{sgbasic} below for key properties of symplectic groupoids.
The notion of a bibundle for symplectic groupoids is an
adaptation of Definition \ref{Gaction}, now also involving the 
idea of a symplectic groupoid action \cite{MiWe}.
\begin{definition}\label{defsgaction}
An action of a symplectic groupoid $\Gm$ on a symplectic manifold
$S$ is called symplectic when the graph of the action in
$\Gm\x S\x S^-$ is Lagrangian.

Let $\Gm,\Sg$ be symplectic groupoids. A (right principal) symplectic $\Gm$-$\Sg$
bibundle consists of a symplectic space $S$ that is a (right principal) bibundle as in
Definition \ref{Gaction}, with the additional requirement that the two
groupoid actions be symplectic.
\end{definition}

The tensor product of two matched right principal bibundles for symplectic groupoids
is then defined exactly as in the general (non-symplectic) case, viz.\ 
by (\ref{btp}). Compared with Lemma \ref{newlemma}, one now needs
the fact that $S_1\otg_{\Sg}S_2$ is symplectic when $S_1$ and $S_2$ are,
and the pertinent actions of $\Sg$ are symplectic. For this, see
Prop.\ 2.1 in \cite{X1}. Also, the notion of isomorphism for symplectic
bibundles is the same as for the Lie case, with extra requirement that
the pertinent diffeomorphism is a symplectomorphism.
Finally, if $G=\Sg$ is a symplectic groupoid, then the canonical
$\Sg$-$\Sg$ bibundle $\Sg$ is symplectic; cf.\ \cite{CDW}.

Hence we may specialize Definition \ref{defLG}
to
\begin{definition}\label{defSyG}
The category \SyG\ has symplectic groupoids as objects, and isomorphism
classes of right principal symplectic bibundles as arrows. The arrows are composed by
the bibundle tensor product (\ref{btp}), for which the canonical bibundles
 $\Sg$ are   units.
\end{definition}
\subsection{Morita equivalence for symplectic groupoids}
We paraphrase Xu's definition of Morita equivalence for symplectic groupoids
\cite{X1}:
\begin{definition}\label{defMESG}
Two symplectic groupoids $\Gm,\Sg$ are called Morita equivalent 
when there exists
a biprincipal  symplectic $\Gm$-$\Sg$ bibundle $S$ 
(called an equivalence symplectic bibundle).
\end{definition}

Cf.\ Definition \ref{MEGR}. The symplectic analogue of Proposition \ref{MEGR2} is
\begin{proposition}\label{MESyG}
A symplectic bibundle $S\in (\Gm,\Sg)$ is an equivalence symplectic bibundle iff
its isomorphism class $[S]$ is invertible
as an arrow  in \SyG.

In other words, two symplectic groupoids  are isomorphic objects in \SyG\
iff they are Morita equivalent.
\end{proposition}
\begin{Proof}
The proof is practically the same as for Proposition \ref{MEGR2}, since
it is already given that $S$ and $S\inv$ are symplectic. The only
difference is that $S\inv$ as a
symplectic manifold should be defined as $S^-$, that is, as $S$ with
minus its symplectic form.
\end{Proof}

The  following definition and proposition are due to Xu \cite{X1,X2}.
\begin{definition}\label{Repsg}
The objects in the \rep\ category $\Rep^s(\Gm)$ of a symplectic groupoid
$\Gm$ are symplectic left $\Gm$ actions on 
smooth maps $\ta:S\raw\Gm_0$, where $S$ is symplectic.
The space of intertwiners is as in Definition \ref{Repgoid}, with
the additional requirement that $\phv$ be a complete Poisson map.
\end{definition}

As in Proposition \ref{Morgr}, we have, with the same proof,
\begin{proposition} \label{Morsg}
If two symplectic groupoids are related by a symplectic equivalence
bibundle, then their \rep\ categories $\Rep^s(\cdot)$ are equivalent.
\end{proposition}
\section{Poisson manifolds}
A Poisson algebra is a commutative associative algebra $A$
(over $\C$ or $\mathbb R$) endowed with a Lie bracket $\{\, ,\, \}$
such that each $f\in A$ defines a derivation $X_f$ on $A$ (as a
commutative algebra) by $X_f(g)=\{f,g\}$. In other words, the Leibniz
rule $\{f,gh\}=\{f,g\} h+g\{f,h\}$ holds.  Poisson algebras are the
classical analogues of \ca s and \vna s; see, e.g.,
\cite{NPL3}.

A Poisson manifold is a manifold $P$ with a Lie bracket on $\cin(P)$
such that the latter becomes a Poisson algebra under pointwise
multiplication. We write $P^-$ for $P$ with minus a given Poisson
bracket.  Not all Poisson algebras are of the form $A=\cin(P)$ (think
of singular reduction), but we specialize to this case, and loosely
think of Poisson manifolds themselves as the classical versions of \ca
s.  The derivation $X_f$ then corresponds to a vector field on $P$,
called the Hamiltonian vector field of $f$. If the span of all $X_f$
(at each point) is $TP$, then $P$ is symplectic.
 General references on symplectic manifolds
are \cite{W0,AM,LM}; for Poisson manifolds see \cite{Vai}.
\subsection{The category \Po\ of Poisson manifolds and dual pairs}
The definition of a suitable category of Poisson manifolds will be based on the notion
of a dual pair. This
 concept, which plays a central role in the interaction between
 symplectic and Poisson geometry, is due to Weinstein \cite{W1} and
 Karasev \cite{K1}; also cf.\
\cite{CDW,NPL3,CW}). Note that these authors all impose somewhat different technical
 conditions.
\begin{definition}\label{defsb}
A  dual pair $Q\law S \raw P$ consists of a
symplectic manifold $S$, Poisson manifolds $Q$ and $P$, and complete Poisson
maps $q:S\raw Q$ and $p:S\raw P^-$, such
that $\{q^* f, p^* g\}=0$ for all $f\in\cin(Q)$ and $g\in\cin(P)$.
\end{definition}

Recall that a Poisson map $J:S\raw P$ is called complete when, for every
$f\in\cin(P)$ with complete Hamiltonian flow, the Hamiltonian
flow of $J^*f$ on $S$ is complete as well (that is, defined for
all times).  Requiring a Poisson map to be complete is a classical
analogue of the self-adjointness condition on a \rep\ of a \ca.

We now turn to a possible  tensor product between
dual pairs $Q\law S_1 \raw P$ and $P\law S_2 \raw R$,
supposed to yield a new dual pair $Q\law S_1 \otc_P S_2 \raw
Q$. One problem is that, contrary to both the purely algebraic
and the $C^*$- algebraic situation, such a tensor product 
does not always exist. To explain the conditions guaranteeing
existence, and also to describe the natural context for this
tensor product, we first recall the notion
of symplectic reduction \cite{LM,W0}.

Let $(\SSS,\om)$ be a symplectic manifold, and let $C$ be a closed
submanifold of $\SSS$.  The null distribution distribution $\CN_C$ on
$C$ is the kernel of the restriction $\om_C=\io^*\om$ of $\om$ to $C$;
here $\io:C\hraw \SSS$ is the canonical embedding.  We denote the
annihilator in $T^*\SSS$ of a subbundle $V\subset T\SSS$ by $V^0$. For
example, $\CN_C^0$ consists of all 1-forms $\al$ on $\SSS$ that
satisfy $\al(X)=0$ for all $X\in \CN_C$.  The symplectic orthogonal
complement in $T\SSS$ of $V$ is called $V^{\perp}$; it consists of all
$Y\in T\SSS$ such that $\om(X,Y)=0$ for all $X\in V$. In this notation
we obviously have $\CN_C=TC\cap TC^{\perp}$.

The following result describes regular symplectic reduction.
\begin{lemma}\label{clredsp}
When the rank of $\om_C$ is constant on $C$, the null distribution
$\CN_C$ is smooth and completely integrable; denote the corresponding
foliation of $C$ by $\Phi_C$.
In addition, assume that the
space $\SSS^C:=C/\Ph_C$  of leaves of this foliation
is a manifold in its natural topology. 

Then there is a unique symplectic
form $\om^C$ on $\SSS^C$ satisfying $\ta_{C\raw \SSS^C}^*\om^C=\om_C$.
\end{lemma}

Here $\ta_{C\raw \SSS^C}$ maps $\sg$ to the leaf of the null
foliation in which it lies. For a proof cf.\ \cite{LM}. 

If one drops either of the assumptions in the proposition, one enters the
domain of singular symplectic reduction, in which it is no longer
guaranteed that the reduced space is a symplectic manifold.
We now specialize to dual pairs.
\begin{lemma}\label{ssr}
Let $Q \law S_1 \raw P$ and
$P \law S_2 \raw R$ be 
dual pairs, with Poisson maps $J_L:S_1\raw P^-$ and $J_R:S_2\raw P$.
Assume that $$T_pP=(T_xJ_L)(T_xS_1)\oplus (T_yJ_R)(T_yS_2)$$ for all
$(x,y)\in S_1 \times_P S_2$, where $p=J_L(x)=J_R(y)$ (for example, it
suffices that either $J_L$ or $J_R$ is a surjective submersion, or,
more weakly, that either $TJ_L$ or $TJ_R$ is surjective at all points
relevant to $S_1 \times_P S_2$).

Then the first assumption in Lemma \ref{clredsp} holds, with
$\SSS=S_1\x S_2$ and $C=S_1 \times_P S_2$. In case that the second
assumption holds as well, one obtains a symplectic manifold \begeq S_1
\otc_P S_2=(S_1 \times_P S_2)/\CN_C \label{defotc}
\end{equation}
and a dual pair
\begeq
Q\law S_1 \otc_P S_2 \raw R.
\end{equation}
\end{lemma}

This lemma is a rephrasing of Thm.\ IV.1.2.2 in \cite{NPL3}, which
in turn is a reformulation of Prop. 2.1 in \cite{X1}. 
\begin{Proof}
The (routine) proof
may be adapted from these references. The maps $q_1:S_1 \otc_P S_2 \raw Q$ and
$r_2:S_1 \otc_P S_2 \raw R^-$ are simply given, in obvious notation,
by $q_1([x,y])=q(x)$ and $r_2([x,y])=r(y)$, where $q:S_1\raw Q$ and
$r:S_2\raw R^-$ are part of the data of the original 
dual pairs; the point is that these maps are well defined as a
consequence of Noether's theorem (in Hamiltonian form \cite{AM,NPL3}).
The same theorem implies the completeness of $q_1$ and $r_2$, for the
Hamiltonian flow of $q^*f$ on $S_1$, $f\in\cin(Q)$, composed with the
trivial flow on $S_2$ so as to lie in $S_1\x S_2$, leaves $S_1 \times_P
S_2$ stable.  Hence the Hamiltonian flow of $q_1^*f$ on $S_1 \otc_P
S_2$ is simply the canonical projection of its flow on $S_1\times_P S_2$,
which is complete by assumption (and analogously for $r$).\end{Proof}

In order to explain which Poisson manifolds and dual pairs
are going to be contained in the category \Po, we invoke the theory
of symplectic groupoids; cf.\ the preceding section.
In the context of Poisson manifolds, we recall the following
features \cite{CDW,MiWe}.
\begin{lemma}\label{sgbasic}
In a symplectic groupoid $\Gm$:
\begin{enumerate}
\item
$\Gm_0$ is a Lagrangian submanifold of $\Gm_1$;
\item The inversion $I$ is an anti-Poisson map;
\item There exists a unique Poisson structure on $\Gm_0$ such that
$t$ is a Poisson map and $s$ is an anti-Poisson map;
\item The foliations of $\Gm$ defined by the
levels of $s$ and $t$ are mutually symplectically orthogonal;
\item If $\Gm$ is s-connected, then $s^*\cin(\Gm_0)$ and
$t^*\cin(\Gm_0)$ are each other's Poisson commutant.
\end{enumerate}
\end{lemma}

The objects in \Po\ are now defined as follows.
\begin{definition}\label{defintP}
A Poisson manifold $P$ is called integrable when there
exists an $s$-connected and $s$-simply connected
symplectic groupoid $\Gm(P)$ over $P$ (so that $P$ is isomorphic
to $\Gm(P)_0$ as a Poisson manifold).
\end{definition}

This definition has been adapted from \cite{CDW}, where no connectedness
requirements are made.
\begin{lemma}\label{unique}
An $s$-connected and $s$-simply connected symplectic groupoid $\Gm(P)$
over an integrable Poisson manifold $P$ is unique up to isomorphism.
\end{lemma}

\begin{Proof} 
We recall that the Lie algebroid of a symplectic groupoid $\Gm$ is
(isomorphic to) $T^*\Gm_0$ \cite{CDW,Vai}.
 Hence $T^*P$ is integrable as a Lie
algebroid when $P$ is integrable as a Poisson manifold, and $\Gm(P)$
is simultaneously the integral of the Lie algebroid $T^*P$ (where
$\Gm(P)$ is seen as a Lie groupoid) and of the Poisson manifold $P$
(where $\Gm(P)$ is seen as a symplectic manifold). Now, Prop.\ 3.3 in
\cite{MM} guarantees that if a Lie algebroid comes from a Lie
groupoid, then the latter may be chosen so as to be s-connected and
s-simply connected; by Prop. 3.5 in \cite{MM}, it is then unique up to
isomorphism. Hence the uniqueness of $\Gm(P)$ as claimed follows from its
uniqueness as the Lie groupoid with Lie algebroid $T^*P$.\end{Proof}

To define the arrows in \Po, we recall a crucial fact
about symplectic groupoid actions.
\begin{lemma}\label{DX}
\begin{enumerate}
\item
The base map $\rh:S\raw \Gm_0$ of a symplectic action of a symplectic
groupoid $\Gm$ on a symplectic manifold $S$  is
a complete Poisson map. Beyond the definition
of a groupoid action, the $\Gm$ action is related to the base
map by the following property. For $(\gm,y)\in \Gm\times^{s,\rh}_{\Gm_0}S$ 
with $\gm=\phv_1^{t^*f}(\rh(y))$,
one has $\gm y=\phv^{\rh^*f}_1(y)$. Here $\phv_t^g$ is the Hamiltonian
flow induced by a function $g$, and $f\in\cin(\Gm_0)$.
\item
Conversely, when $\Gm$ is s-connected and
s-simply connected, a given complete Poisson map $\rh:S\raw \Gm_0$
is the base map of a unique symplectic $\Gm$ action on
$S$ with the above property.
\end{enumerate}
\end{lemma}

The first claim is taken from \cite{MiWe,CDW}, and the second is due to 
\cite{Daz,X2}. 
\begin{lemma}\label{4lemma}
\begin{enumerate}
\item
Let $P$ and $Q$ be integrable Poisson manifolds, with associated
s-connected and s-simply connected symplectic groupoids $\Gm(P)$ and
$\Gm(Q)$; cf.\ Definition \ref{defintP}. 
There is a natural bijective correspondence between
dual pairs $Q\law S\raw P$ and symplectic bibundles
$\Gm(Q)\rac S \lac\Gm(P)$.
\item In particular, the dual pair associated to the
canonical bibundle $\Gm(P)\rac\Gm(P)\lac\Gm(P)$
is $P\stackrel{t}{\law}\Gm(P)\stackrel{s}{\raw}P$.
\item
Let $R$ be a third integrable Poisson manifold, with associated
s-connected and s-simply connected symplectic groupoid $\Gm(R)$,
and let $Q\law S_1\raw P$ and 
$P\law S_2\raw R$ be dual pairs. In case that
the associated symplectic bibundles are right principal, one has
\begeq
S_1\otc_{P} S_2=S_1\circledast_{\Gm(P)} S_2 \label{rhsxu}
\end{equation}
as symplectic manifolds, as $\Gm(Q)$-$\Gm(R)$ symplectic bibundles, and as
 $Q$-$R$ dual pairs.
\end{enumerate}
\end{lemma}

\begin{Proof} The first claim follows from Lemma \ref{DX} and the Hamiltonian
Noether theorem, in the form that states that $[\phv_s^f,\phv_t^g]=0$
for all times $s,t$ iff $\{f,g\}=0$ (provided that $f$ and $g$ have complete
Hamiltonian flows) \cite{AM,LM}. 

The second claim is immediate from 
Definition \ref{defintP} and Lemma \ref{sgbasic}.3 and 5.

The third claim is a rephrasing of the proof of Prop.\ 2.1 in \cite{X1}.
\end{Proof}

We say that two $Q$-$P$ dual pairs
$Q\stackrel{q_i}{\law}\til{S}_i\stackrel{p_i}{\raw}P$, $i=1,2$,
are isomorphic when there is a symplectomorphism 
$\phv:\til{S}_1\raw \til{S}_2$ for which $q_2\phv=q_1$ and $p_2\phv=p_1$.
This squares with Lemma \ref{DX}, in that it is compatible with
the notion of isomorphism between symplectic bibundles (defined after
Definition \ref{defsgaction}). In other words, the bijective correspondence
 between complete Poisson
maps and symplectic groupoid actions behaves naturally under isomorphisms.
\begin{lemma}
\begin{enumerate}
\item
One has isomorphisms $S_1\otc_P\Gm(P)\simeq S_1$ and $\Gm(P)\otc_P S_2\simeq S_2$
as $Q$-$P$ and $P$-$R$ dual pairs, respectively.
\item The tensor product $\otc$ is associative up to isomorphism, and passes
to isomorphism classes of dual pairs.
\end{enumerate}
\end{lemma}

\begin{Proof}
The first claim follows from  (\ref{rhsxu})  and Lemma \ref{GGlem}.
  Alternatively, it may
be established  by direct calculation: for example, the symplectomorphism
$\Gm(P)\otc_P S\raw S$ is given by $[\gm,y]\mapsto \gm y$.

The second claim follows from (\ref{rhsxu}) and
the last item in Lemma \ref{newlemma}, or, alternatively, directly from
Lemma \ref{ssr}.
\end{Proof}

We are now, at last, in a position to give a classical version of Definition
\ref{defCa}.
\begin{definition}\label{defPo}
We say that a dual pair $P\law S\raw Q$ between integrable
Poisson manifolds is regular when the associated symplectic bibundle
$\Gm(P)\rac S\lac\Gm(Q)$ is right principal (cf.\ Definition \ref{defprin}
and Lemma \ref{DX}).

The category \Po\ has integrable Poisson manifolds as objects, and isomorphism
classes of regular dual pairs as arrows. The arrows are composed by
the tensor product $\otc$ (cf.\ (\ref{defotc})), for which the dual pairs
$P\stackrel{t}{\law}\Gm(P)\stackrel{s}{\raw}P$ are   units (cf.\ Lemma \ref{4lemma}.2).
\end{definition}

It is clear from Lemma \ref{4lemma} that \Po\ is equivalent to the full
subcategory of \SyG\ whose objects are s-connected and s-simply connected
symplectic groupoids.
\subsection{Morita equivalence for Poisson manifolds}
The theory of Morita equivalence of Poisson manifolds was initiated by
Xu \cite{X2},
 who gave the following definition. 
\begin{definition}\label{MEPM}
A dual pair $Q\law S\raw P$ is called an equivalence
 dual pair when:
\begin{enumerate}
\item
The maps $p:S\raw P$ and $q:S\raw Q$ are surjective submersions;
\item The level sets of $p$ and $q$ are connected and simply connected;
\item The foliations of $S$ defined by the
levels of $p$ and $q$ are mutually symplectically orthogonal (in that
the tangent bundles to these foliations are each other's
  symplectic orthogonal complement).
\end{enumerate}
Two Poisson manifolds related by an equivalence dual pair are called Morita equivalent.
\end{definition}

This definition enables us to reappreciate the definition of integrability.
\begin{lemma}
A Poisson manifold is integrable iff it is Morita equivalent to itself.
\end{lemma}

\begin{Proof} 
As remarked in \cite{X2}, the proof of ``$\Law$'' is Cor.\ 5.3 in 
\cite{W2}.

The ``$\Raw$'' claim follows because one may take $S=\Gm(P)$
in Definition \ref{MEPM};
condition 1 in  is satisfied by definition of a
symplectic groupoid, condition 2 follows by assumption, and condition
3 is proved in section II.1 of \cite{CDW} (Corollaire following
Remarque 2) or in Thm.\ 1.6 of \cite{MiWe}.  
\end{Proof}

We have now arrived at the desired conclusion:
\begin{proposition}\label{MEPo}
A regular symplectic bimodule $S\in (P,Q)$ is invertible 
as an arrow in \Po\ 
iff it is an equivalence symplectic bimodule.

In other words, two integrable Poisson manifolds  are isomorphic objects in \Po\
iff they are Morita equivalent.
\end{proposition}
\begin{Proof}
The proof will be based on the following lemma, which, of course, is of
great interest in itself.
\begin{lemma}\label{xu3.2}
Let $P$ and $Q$ be integrable Poisson manifolds, with s-connected and
s-simply connected symplectic groupoids $\Gm(P)$ and $\Gm(Q)$.
 Then $P$ and $Q$ are \Me\ as Poisson manifolds iff $\Gm(P)$ and $\Gm(Q)$
are \Me\ as symplectic groupoids.
\end{lemma}

This is Thm.\ 3.2 in \cite{X2}, to which we refer for the proof.
Now, by Proposition \ref{MESyG}, $\Gm(P)$
and $\Gm(Q)$ are Morita equivalent iff
$\Gm(Q)\cong\Gm(P)$ in \SyG, 
which is true iff there is an invertible symplectic
bibundle $\Gm(Q)\rac S\lac\Gm(P)$ in \SyG, with inverse
$\Gm(P)\rac S^-\lac\Gm(Q)$. 
Hence
\begin{eqnarray}
\Gm(P)\rac S^- \otg_{\Gm(Q)} S\lac \Gm(P) & \simeq  & \Gm(P)\rac \Gm(P)\lac \Gm(P);  \label{SyM1}\\
\Gm(Q)\rac S\otg_{\Gm(P)} S^- \lac \Gm(Q) & \simeq & \Gm(Q)\rac \Gm(Q)\lac \Gm(Q). \label{SyM2}
\end{eqnarray}

By Lemma \ref{4lemma}.1 and 3, and the compatibility of isomorphisms
for symplectic bibundles and their associated symplectic bimodules,
this is equivalent to
\begin{eqnarray}
P\law S^- \otc_Q S\raw P & \simeq  & P\law \Gm(P)\raw P;  \label{PoM1}\\
Q\law S\otc_P S^- \raw Q & \simeq & Q\law \Gm(Q)\raw Q. \label{PoM2}
\end{eqnarray}

By Lemma \ref{xu3.2}, this means that $Q\cong P$ in \Po.
\end{Proof}

As a corollary, note that an equivalence symplectic bimodule is automatically
regular (since an equivalence symplectic bimodule is regular by Proposition \ref{MESyG}).

The following definition of the \rep\ category of a Poisson manifold is
simpler than the one used in \cite{X1,X2}, but leads to the same
Morita theorem.
\begin{definition}\label{MEPM2}
The \rep\ category $\Rep(P)$ of a Poisson manifold has complete
Poisson maps $J:S\raw P$, where $S$ is some symplectic space, as
objects, and complete Poisson maps $\phv:S_1\raw S_2$, where
$J_2\phv=J_1$, as arrows.
\end{definition}

One then has Xu's Morita theorem for Poisson manifolds \cite{X2}:
\begin{proposition}\label{xuthm}
If two integrable Poisson manifolds are Morita equivalent,
then their \rep\ categories are equivalent.
\end{proposition}

This is proved as for algebras.
Xu's proof was based on
 the following extraordinary property,
described locally in \cite{CDW}, and globally in
\cite{Daz,X2}. 
\begin{proposition}\label{xd}
If $\Gm$ is an $s$-connected and $s$-simply connected
symplectic group\-o\-id, with associated Poisson manifold
$\Gm_0$ (cf.\ Proposition \ref{sgbasic}.3), then
the \rep\ categories $\Rep^s(\Gm)$ and $\Rep(\Gm_0)$ are equivalent.
\end{proposition}
\begin{Proof}
This is immediate from Lemma \ref{DX}.
\end{Proof}

Proposition \ref{xuthm} now
trivially follows from Lemma \ref{xu3.2} and Propositions \ref{Morsg} and
 \ref{xd}. This was Xu's original argument \cite{X2}. 

\section{Marsden--Weinstein--Meyer reduction}
The (regular) Marsden--Weinstein--Meyer reduction procedure in symplectic geometry 
(see \cite{W0,AM,LM} for the usual theory) may be reformulated as 
a special case of Lemma \ref{ssr}. This will be explained in the first
section below. In the subsequent two sections we will write down
 analogous reduction processes for \ca s and \vna s, which should be seen
as quantized versions of Marsden--Weinstein--Meyer reduction; cf.\ the Introduction.
\subsection{Classical Marsden--Weinstein--Meyer reduction}
In Lemma \ref{ssr} we take $S_1=S$ to be a symplectic manifold
equipped with a strongly Hamiltonian action of a connected Lie group
$H$; hence there exists a Poisson map $J:S\raw (\h^*)^-$, where
$\h^*$, the dual of the Lie algebra $\h$ of $H$, is equipped with the
Lie--Poisson structure.  This map is automatically $H$-equivariant
with respect to the coadjoint $H$-action on $\h^*$.  We now take
$P=\h^*$ and $J_L=J$. In case that the $H$-action is free and proper,
the quotient $Q=S/H$ inherits the Poisson structure from $S$, and
thereby becomes a Poisson manifold (which in general fails to be
symplectic). The canonical projection $q$ is a Poisson
map. Furthermore, we take $S_2=0$ and $J_R$ to be the embedding of 0
into $\h^*$. Finally, $R$ is a point.  Thus the two 
dual pairs in Lemma \ref{ssr} are taken to be $S/H\law S\raw\h^*$ and
$\h^*\law 0\raw pt$.  For the completeness of the pertinent maps, see
\cite{X3}, or Prop.\ IV.1.5.8 in \cite{NPL3}.
 
It now follows from direct computation, or from the general theory of
symplectic reduction, that the classical tensor
product of these
 dual pairs is \begeq 
S/H\law S\circledcirc_{\h^*}0\raw pt\:\:\simeq\:\:
S/H \law J\inv (0)/H\raw pt,
\end{equation}
with the obvious maps. 
(In case that $H$ is disconnected one would take the quotients by the
connected component of the identity.) For this space to be a
symplectic manifold, it actually suffices that $H$ acts freely and
properly on $J\inv(0)$; this is, of course, no guarantee that $S/H$ is
a manifold. The singular case has been extensively studied in the
``intermediate'' case in which the $H$-action is proper but not free;
see the present volume.

An apparent generalization would be to take $S_2$ to be a coadjoint
orbit $\CO$ in $\h^*$, endowed with the Lie--Kirillov--Kostant--Souriau
 symplectic
structure; in that case the embedding $\iota$ is a Poisson map.  One
then has
\begeq 
S\otc_{\h^*}\CO\simeq J\inv(\CO)/H. \label{kks}
\end{equation}

This is not really a generalization of the case where the orbit is 0,
since in the latter case one may always replace $S$ by $S\x\CO^-$, on
which $H$ acts by the product of the given action on $S$ and the
coadjoint action on $\CO^-$. The momentum map $J_{S\x\CO^-}$
is then the sum of the original one $J=J_S$ on $S$, and minus the embedding
map $\iota_{\CO}:\CO\hookrightarrow\h^*$, i.e., 
\begeq
J_{S\x\CO^-}=J_S-\iota_{\CO}.\label{shift}
\end{equation}
 Hence one obtains the same reduced space $J\inv(\CO)/H$
(the ``shifting trick'').

In any case, for arbitrary $\CO$ and
$S=T^*G$, where $G$ is a Lie group containing $H$ as a closed subgroup
(which acts on $T^*G$ by lifting its natural right action on $G$), 
the classical tensor product yields the symplectic spaces studied by
Kazhdan--Kostant--Sternberg \cite{KKS}. These were introduced in order
to mimic Mackey's induced representations in a classical setting.

The classical tensor product also covers the case of reduction with
respect to a momentum map $J:S\raw\h^*$ that is not equivariant with
respect to the coadjoint action $\Co$ on $\h^*$. In that case one
proceeds as follows \cite{LM,NPL3}. First, compute the so-called
symplectic cocycle $\gm$ on $H$ with values in $\h^*$, given by
\begeq
\gm(h)=J(h\sg)-\Co(h)J(\sg),
\end{equation}
which turns out to be independent of $\sg\in S$. Next, define a
 2-cocycle $\Gm$ on $\h$ by
\begeq
 \Gm(X,Y) =-
\frac{d}{dt}\gm(\Exp(tX))(Y)_{|t=0}. \label{defGm1}
\end{equation}
This leads to a
modified Poisson bracket on
 $\h^*$, given 
 by 
\begeq
 \{f,g\}_{\Gm}=\{f,g\}+ \Gm(df,dg).
 \ll{modliepo} 
\end{equation}
We denote $\h^*$ with this Poisson structure by $\h^*_{\Gm}$; the
momentum map $J:S\raw (\h^*_{\Gm})^-$ is Poisson.  The symplectic
leaves of $\h^*_{\Gm}$ are the orbits of the twisted coadjoint action
$\Co_{\gm}$ of $H$ on $\h^*$, given by
\begeq
\Co_{\gm}(h)\theta=\Co(h)\theta+\gm(h).\label{tca}
\end{equation}
Being symplectic leaves, these orbits acquire a symplectic structure.

One then picks a $\Co_{\gm}(H)$ orbit $\CO_{\gm}\subset\h^*$, and
takes the  dual pairs $S/H\law S\raw\h^*_{\Gm}$ and
$\h^*_{\Gm}\law \CO_{\gm}\raw pt$, as before. The classical tensor
product becomes 
\begeq 
S\otc_{\h^*_{\Gm}}\CO_{\gm}\simeq J\inv(\CO_{\gm})/H. \label{kks2}
\end{equation}

Despite the formal similarity between (\ref{kks2}) and (\ref{kks}),
the reduced spaces in the equivariant and the non-equivariant cases
tend to be vastly different. 

One may, alternatively, describe this procedure using the shifting
trick; this elucidates the connection between noneq\-ui\-variant
Marsden--Weinstein--Meyer reduction and the treatment of projective group
\rep s at the end of sections \ref{CMW} and \ref{WMW}.
Namely, we let $H$ act on $\CO^-_{\gm}$ by (\ref{tca}), with momentum
map $J_{\CO^-_{\gm}}=-\iota_{\CO_{\gm}}$.  This momentum map obviously
has symplectic cocycle $-\gm$, so that the momentum map
$J_{S\x\CO_{\gm}^-}=J_S-\iota_{\CO_{\gm}}$ for the $H$-action on
$S\x\CO_{\gm}^-$ (cf.\ (\ref{shift})) has symplectic cocycle 0,
and hence is $\Co$-equivariant. For the reduced space one then has
\begeq
S\otc_{\h^*_{\Gm}}\CO_{\gm}\simeq J_{S\x\CO_{\gm}^-}\inv(0)/H;
\end{equation}
compare Proposition \ref{rindcgr} below.
Thus Marsden--Weinstein--Meyer reduction for a nonequivariant momentum map
amounts to cancelling the pertinent symplectic cocycle by enlarging
the space $S$ to $S\x \CO^-_{\gm}$, much as forming Rieffel's or Connes's 
tensor product  from a projective unitary
\rep\ $U$ on $\H$  necessitates enlarging $\H$ to $\H\ot\H_{\ch}$,
on which $H$ acts without multiplier; see below.
\subsection{$C^*$ Marsden--Weinstein--Meyer reduction}\label{CMW}
 Recall (see, e.g., \cite{NPL3}) that there is a bijective
 correspondence between unitary representations $U_{\ch}$ of a locally
 compact group $H$ and nondegenerate representations $\pi_{\ch}$ of
 its group
\ca\ $C^*(H)$, given by
\begeq
\pi_{\ch}(f)=\int_H dh\, U_{\ch}(h) f(h), \label{pichi}
\end{equation}
where $f\in L^1(H)$, $dh$ is the Haar measure, and for simplicity we
have assumed that $H$ is unimodular.  In particular, a unitary
representation $U_{\ch}(H)$ on a Hilbert space $\H_{\ch}$ yields a
Hilbert bimodule $C^*(H)\rac\H_{\ch}\rlh\C$.  A slight modification of
(\ref{pichi}) associates an anti-representation (or right action) of
$C^*(H)$ to a unitary representation of $H$; see (\ref{rmw3.5}) below.

$C^*$  Marsden--Weinstein--Meyer reduction is Rieffel's interior tensor product
of the Hilbert bimodules $\BH^H\rac \H^-\rlh C^*(H)$ and
$C^*(H)\rac\H_{\ch}\rlh\C$.  Here $H$ is a Lie group, and $\H$ is a
Hilbert space carrying a unitary representation $U$ of $H$; the space
$\H^-$ is a completion (different from $\H$) of a certain dense subspace of
$\H$. Furthermore, $\BH^H$ is the \ca\ of $H$-invariant bounded
operators on $\H$ (that is, the commutant of $U(H)$), and $\H_{\ch}$
is a second Hilbert space carrying a unitary representation of $H$
(often the trivial one).

For simplicity we initially assume that $H$ is compact (hence unimodular).
The dense subspace mentioned above is then simply $\H$ itself.
For the notion of a pre-Hilbert $C^*$-module
and its completion occurring below see \cite{Rie1}, \cite{RW}, or \cite{NPL3}. 
\begin{proposition}\ll{1com}
Let $U$ be a \rep\ of a compact Lie group $H$ on a \Hs\ $\H$, with
corresponding \rep\ $\pi$ of the group $C^*$-algebra $C^*(H)$.  The
formula
\begeq \pi_R(f)=\int_H dh\, f(h)U(h)^{-1}
\label{rmw3.5} \end{equation}
 defines a right action $\pi_R$ of $C^*(H)$ by
continuous extension from $f\in\cci(H)$. In conjunction with the map
$\la\, ,\,\ra_{C^*(H)}:\H\x\H\raw C^*(H)$, defined by 
\begeq
 \la
\Ps,\Ph\ra_{C^*(H)}:\, h\mapsto (\Ps,U(h)\Ph), \label{rmw3.6} 
\end{equation}
one obtains a pre-Hilbert $C^*$-module $\H\rlh C^*(H)$.  Completion
produces a Hilbert bimodule  $\BH^H\rac\H^-\rlh C^*(H)$.
\end{proposition}

For the proof cf.\ \cite{NPL3}, IV.2.5. 

It is easy to describe the interior tensor product of
$\BH^H\rac\H^-\rlh C^*(H)$ and $C^*(H)\rac\H_{\ch}\rlh\C$.
\begin{proposition}\ll{rindcgr}
For compact $H$, the Hilbert bimodule
$$\BH^H\rac\H^-\hat{\ot}_{C^*(H)}\H_{\ch}\rlh\C$$ is isomorphic to
$$\BH^H_0\rac (\H\ot\Hlg)_0\rlh\C,$$
where $(\H\ot\Hlg)_0$ is
the invariant subspace of $\H\ot\Hlg$ under 
$U\ot U_{\ch}(H)$, and $\BH^H_0$ is the
restriction of $\BH^H$ to $(\H\ot\Hlg)_0$.
\end{proposition}

This follows from a straightforward computation, and coincides with what
physicists have known since the work of Dirac on constrained quantization.

The more interesting case where $H$ is not compact, and possibly not
even locally compact, is discussed in detail in \cite{NPL3}.
The special case $\H=L^2(G)$ and $H\subset G$, acting on $\H$ in the
right-regular representation, was already discussed by Rieffel \cite{Rie1}.

In summary, one first needs to find a dense subspace $\CD\subset \H$ such that
the expression $\int_H dh\, (\Ps,U(h)\Ph)$ is finite for all $\Ps,\Ph\in\CD$.
For example, for $\H=L^2(G)$ one may take $\CD=\cci(G)$.
The space $\H^-$ is then the pertinent completion of $\CD$ 
as a Hilbert $C^*$ module, rather than
 $\H$. The Hilbert space $\H^-\hat{\ot}_{C^*(H)}\H_{\ch}$
is formed by first endowing $\CD\ot_{\C}\H_{\ch}$ with the 
sesquilinear form
\begeq
(\til{\Ps},\til{\Ph})_0=\int_H dh\, ( \til{\Ps},U\ot
U_{\ch}(h)\til{\Ph})_{\H\ot\Hlg},\ll{fundrig} 
\end{equation}
which is positive semidefinite. One then takes the quotient
$(\CD\ot_{\C}\H_{\ch})/\CN_0$, where $\CN_0$ is the null space of $(\,
,\, )_0$. This quotient inherits the latter form, which is now a
(positive definite) inner product. The completion of
$(\CD\ot_{\C}\H_{\ch})/\CN_0$ in the norm derived from the inner product
is $\H^-\hat{\ot}_{C^*(H)}\H_{\ch}$. 
The \ca\ $\A=\BH^H$ for compact $H$ now needs to be replaced by a suitable
dense subalgebra whose elements leave $\CD$ stable.  Such elements
$A$ act on $\H^-\hat{\ot}_{C^*(H)}\H_{\ch}$ in the obvious way, by
quotienting $A\ot\I_{\ch}$.

The above theory may be generalized to projective
unitary \rep s. Let $c$ be a multiplier on $H$, taking values
in $U(1)$. This leads to the twisted group \ca\ $C^*(H,c)$;
see \cite{NPL3} and refs.\ therein. Eq.\ (\ref{pichi}) now
establishes a bijective correspondence between nondegenerate
\rep s of $C^*(H,c)$ and projective unitary \rep s with multiplier
$c$; that is, one has $U_{\ch}(x)U_{\ch}(y)=c(x,y)U_{\ch}(xy)$.

 Now suppose that $U(H)$ is a projective unitary \rep\ with multiplier
 $\ovl{c}$.  Using precisely the same formulae as in the nonprojective
 case, Proposition \ref{1com} remains valid with $C^*(H,c)$
 replacing $C^*(H)$, so that one obtains a Hilbert bimodule
 $\BH^H\rac\H^-\rlh C^*(H,c)$. Similarly, a projective unitary \rep\
 $U_{\ch}$ with multiplier $c$ yields a Hilbert bimodule
 $C^*(H,c)\rac\H_{\ch}\rlh\C$. In the compact case the interior  tensor
 product of these Hilbert bimodules remains described by Proposition
 \ref{rindcgr}; the concept of an invariant subspace
of $\H\ot\H_{\ch}$ under $U\ot U_{\ch}(H)$ continues to make sense,
since the latter \rep\ is no longer projective.

\subsection{$W^*$ Marsden--Weinstein--Meyer reduction}\label{WMW}
 A $W^*$-analogue of the $C^*$ reduction procedure of the preceding
 section suggests itself, since the \ca\ $\BH^H$ occurring there is
 automatically a \vna.  The \vna\ of $H$, denoted by $W^*(H)$, is the
 weak closure of $C^*_r(H)$, which in turn is the regular
\rep\ of the group \ca\ $C^*(H)$ on $L^2(H)$; cf.\ \cite{NPL3}.

Any continuous unitary \rep\ $U_{\ch}$ of $H$ that is weakly contained in the
regular one defines a \rep\ $\pi_{\ch}$ of $C^*_r(H)$ by (\ref{pichi}),
which extends to a \rep\ of $W^*(H)$ denoted by the same symbol.
Hence if both $U$ and $U_{\ch}$ are weakly contained in the regular \rep,
one has correspondences $\BH^H\rac\H\lac W^*(H)$ and
$W^*(H)\rac\H_{\ch}\lac\C$, whose relative tensor product
$ \BH^H\rac \H\boxtimes_{W^*(H)}\H_{\ch}\lac\C$
is a quantized, \vna ic version of Marsden--Weinstein--Meyer reduction
 in its own right. 

Under the standing assumption that the pertinent \rep s are
weakly contained in the
regular one, we now examine the relationship between $W^*$ and $C^*$
Marsden--Weinstein--Meyer reduction.
\begin{proposition}\label{intcf}
Define the Hilbert bimodule $\BH^H\rac\H^-\rlh
C_r^*(H)$ as in Proposition
\ref{1com}.
 Then the induced \rep s of
$\BH^H$ on $\H^-\hat{\ot}_{C_r^*(H)}\H_{\ch}$ and 
 on $\H\boxtimes_{W^*(H)}\H_{\ch}$ are unitarily equivalent.
\end{proposition}

Note that these \rep s may either be seen 
as  $\BH^H$-$\C$ Hilbert bimodules or as $\BH^H$-$\C$ correspondences;
cf.\ Example \ref{BDHex}.1.
\begin{Proof}
The proof is based on the well-known fact that
$W^*(H)\rac L^2(H)$ is in standard form. Using Lemma \ref{CSlem},
we therefore have
\begin{equation}
\H^-\hat{\ot}_{C_r^*(H)}\H_{\ch}\simeq \H^-\hat{\ot}_{C_r^*(H)}(L^2(H)\boxtimes_{W^*(H)}
\H_{\ch}).\label{Heq0}
\end{equation}
In addition, one has
\begin{equation}
\H^-\hat{\ot}_{C_r^*(H)}L^2(H)\simeq   \H. \label{Heq}
\end{equation}
To prove this, note that for $\ps\in\CD$ (defined in the preceding section)
and $f\in\cci(H)$ one has
\begin{equation}
(\ps\ot f,\phv\ot g)_0=(\pi_R(f)\ps,\pi_R(g)\phv)_{\H}, \label{phvHeq}
\end{equation}
where the left-hand side has been defined in (\ref{fundrig}), and
$\pi_R$ is given in (\ref{rmw3.5}).
Since $\pi_R$ is nondegenerate (because of the unitarity of $U$), the map
$\ps\ot f\mapsto \pi_R(f)\ps$ from $\CD\ot\cci(H)$ to $\H$
has dense image, so by (\ref{phvHeq}) it
quotients and extends to a unitary map from $\H^-\hat{\ot}_{C_r^*(H)}L^2(H)$
to $\H$. Combining (\ref{Heq0}) and (\ref{Heq}), noting that all
isomorphisms intertwine the given $\BH^H$-action, and  using the associativity
up to isomorphism of the various tensor products, the proposition follows.
\end{Proof}

The comments on $C^*$ Marsden--Weinstein--Meyer reduction at the end of
section \ref{CMW} now have an evident $W^*$ analogue. 
A projective unitary \rep\ $U(H)$ with multiplier
 $\ovl{c}$ on $\H$ 
defines a correspondence $\BH^H\rac\H\lac W^*(H,c)$,
and similarly a projective unitary \rep\ $U(H)$ with multiplier
 $c$ on $\H_{\ch}$ 
defines a correspondence $W^*(H,c)\rac \H_{\ch}\lac\C$.
Their relative tensor product defines a \rep\ of
$\BH^H$ on $\H\boxtimes_{W^*(H,c)}\H_{\ch}$.
For compact $H$ this \rep\ is described by Proposition \ref{rindcgr}.


\begin{thebibliography}{99}
\itemsep=\smallskipamount

\bibitem{AM} R. Abraham  and
J.E. Marsden, \textit{Foundations of Mechanics}, 2nd ed., Addison Wesley,
Redwood City, 1985.


\bibitem{BDH}
M. Baillet, Y. Denizeau, and J.-F. Havet, \textit{ Indice d'une
esp\'{e}rance conditionnelle},   Compositio Math.\ \textbf{66}
 (1988), 199--236.

\bibitem{Bee} 
W. Beer, \textit{
On Morita equivalence of nuclear $C^{*}$-algebras},
  J.\ Pure Appl.\ Algebra \textbf{26} (1982),  249--267. 

\bibitem{Ble2} D. Blecher, \textit{ On Morita's fundamental theorem
for \ca s}, e-print {\tt math.OA/9906082}.

 \bibitem{BGR} L.G. Brown,  P. Green, and
M.A. Rieffel, \textit{Stable isomorphism and strong Morita equivalence
of $C^*$-algebras}, Pac.\ J.\ Math.\ {\bf 71} (1977), 349--363.

\bibitem{CW} A. Cannas da Silva and A. Weinstein, \textit{Geometric models
for noncommutative geometry}, Berkeley Mathematics Lecture Notes \textbf{10},
American Mathematical Society, Providence, RI, 1999.

\bibitem{Con0} A. Connes, \textit{
On the spatial theory of von Neumann algebras},
J.\ Funct.\ Anal.\ \textbf{35} (1980),  153--164. 

\bibitem{Conmg} A. Connes, \textit{
 Sur la th\'{e}orie non commutative de l'int\'{e}gration,}  in:
 \textit{Alg\`{e}bres
d'op\'{e}rateurs}, pp. 19--143, Lecture
Notes in Math.\ \textbf{725}, Springer, Berlin, 1979.

\bibitem{Con} A. Connes, \textit{Noncommutative Geometry},
Academic Press, San Diego, 1994.

\bibitem{CT} A. Connes and M. Takesaki,  \textit{The flow of weights on
factors of type III}, Tohoku Math. \ J.\  \textbf{29} (1977),  473--575;
Err.\ ibid.\  \textbf{30} (1978),  653--655. 

\bibitem{CDW} A.
Coste, P. Dazord, and A. Weinstein, \textit{
Groupoides symplectiques},  Publ.\ D\'{e}pt.\ Math.\ Univ. C.\
Bernard--Lyon I \textbf{2A} (1987), 1--62.

\bibitem{Daz} P. Dazord,  \textit{Groupo\"{\i}des symplectiques et troisi\`{e}me
 th\'{e}or\`{e}me de Lie ``non lin\'{e}aire}'',  Lecture Notes in
Math.\ \textbf{1416} (1990), 39--74.

\bibitem{faith} C. Faith, \textit{
Algebra: Rings, Modules and Categories.\ I},
 Springer, New York, 1973. 

\bibitem{FHM} J. Feldman,  P. Hahn,  and C.C.  Moore, \textit{Orbit structure and
countable sections for actions of continuous groups},  Adv.\   Math.\  \textbf{28} (1978),
186--230.

\bibitem{GZ} P. Gabriel and M.  Zisman, \textit{Calculus of Fractions and Homotopy Theory},
Springer,  New York, 1967.


\bibitem{Hae} A.
Haefliger, \textit{Groupo\"{\i}des d'holonomie et classifiants},
Ast\'{e}risque \textbf{116} (1984), 70--97.

\bibitem{Hah1} P.  Hahn, \textit{Haar measure for measure groupoids},
 Trans.\ Amer.\ Math.\
Soc.\ \textbf{242} (1978), 1--33.

\bibitem{Hah2} P.  Hahn, \textit{
The regular representations of measure groupoids}, Trans.\ Amer.\ Math.\
Soc. \ \textbf{242} (1978), 35--72.

\bibitem{HS}
M. Hilsum and G. Skandalis, \textit{
 Morphismes $K$-orient\'{e}s d'espaces de feuilles et fonctorialit\'{e}é
 en th\'{e}orie de
Kasparov (d'apr\'{e}s une conjecture d'A. Connes)}. Ann.\ Sci.\ É\'{E}cole 
Norm.\ Sup.\ (4) \textbf{20} (1987),  325--390.

\bibitem{KR1} R.V.
Kadison and J.R. Ringrose, \textit{Fundamentals of the Theory of
 Operator Algebras I. Elementary Theory},  Academic Press, New
 York, 1983.

\bibitem{KR2} R.V.
Kadison and J.R. Ringrose, \textit{Fundamentals of the Theory of
 Operator Algebras II. Advanced Theory},  Academic Press, New
 York, 1986.

\bibitem{K0} M.V. Karasev, \textit{Analogues of objects of Lie group
theory for nonlinear Poisson brackets}, Math.\ USSR Izv.\ \textbf{28}
(1987), 497--527.

\bibitem{K1} M.V.
Karasev, \textit{
 The Maslov quantization conditions in higher cohomology and analogs
 of notions developed in Lie theory for canonical fibre bundles of
 symplectic manifolds.  I, II.},  Selecta Math. Soviet.\ \textbf{8}
 (1989), 213--234, 235--258.

\bibitem{KM} M.V. Karasev and V.P. Maslov, \textit{
Nonlinear Poisson Brackets. Geometry and Quantization},
Translations of Mathematical Monographs, \textbf{119},
 American Mathematical Society,
Providence, RI, 1993.

 \bibitem{KKS} D. Kazhdan,  B. Kostant, and S. Sternberg, \textit{
 Hamiltonian group actions and dynamical systems of Calogero type},
  Commun.\ Pure Appl.\ Math.\  \textbf{31} (1978), 481--507.

\bibitem{vdlaan} P. van der Laan, \textit{Smooth Groupoids}, Univ.\ of Utrecht Lecture
Notes, based on lectures by I. Moerdijk and J. Mr\v{c}un (preprint, 2000).

\bibitem{Lance} E.C.
Lance,  \textit{Hilbert
 $C^*$-Modules. A Toolkit for Operator Algebraists},
  Cambridge University Press, Cambridge, 1995.

\bibitem{NPL1} N.P. Landsman,  
\textit{Rieffel induction as generalized quantum Marsden-Weinstein reduction},
J.\ Geom.\ Phys.\ \textbf{15} (1995), 285--319, and e-print
\texttt{hep-th/9305088}; Err.\ ibid.\ \textbf{17} (1995), 298.

\bibitem{NPL3} N.P. Landsman,  \textit{Mathematical Topics Between Classical
and Quantum Mechanics},
Springer, New York, 1998.

\bibitem{NPLDR} N.P. Landsman,  \textit{Bicategories of operator algebras
and Poisson manifolds}, in \textit{Mathematical Physics in Mathematics
and Physics.  Quantum and Operator Algebraic Aspects}, ed.\ R. Longo,
\textit{Fields Inst.\ Comm.}, to appear (2001), and
e-print \texttt{math-ph/0008003}.

\bibitem{LauNis} R. Lauter and V. Nistor, 
\textit{Analysis of geometric operator on open manifolds: a groupoid
approach}, this volume.

\bibitem{LM} P. Libermann and
C.-M. Marle, \textit{Symplectic Geometry and Analytical Mechanics},
Reidel, Dordrecht, 1987.


\bibitem{Mac} K.C.H.
Mackenzie, \textit{Lie Groupoids and Lie Algebroids in
Differential Geometry},  Cambridge University Press, Cambridge, 1987.

\bibitem{Mackey}  G.W.  Mackey, 
\textit{Ergodic theory and virtual groups},
Math.\ Ann.\ \textbf{166} (1966), 187--207. 


\bibitem{MacLane} S.
Mac Lane, \textit{Categories for the Working Mathematician},
 2nd ed.,  Springer, New York, 1998.

\bibitem{MiWe} K. Mikami  and A. Weinstein, \textit{Moments
and reduction for symplectic groupoids},   Publ.\ RIMS Kyoto Univ.\
\textbf{24} (1988), 121--140. 

\bibitem{Moe} I.
 Moerdijk,   \textit{Toposes and groupoids}, in \textit{Categorical Algebra and its
Applications},  pp.\  280--298, Lecture Notes in Math.\ \textbf{1348},
Springer, Berlin, 1988. 

\bibitem{MM} I.
 Moerdijk and J. Mr\v{c}un, \textit{On integrability of infinitesimal
actions}, Univ.\ of Utrecht preprint (April 2000).

\bibitem{MoS}  C.C. Moore, and C. Schochet, 
\textit{Global Analysis on Foliated Spaces}, 
Springer, New York, 1988.

\bibitem{Mrc1}  J. Mr\v{c}un, \textit{Stability and Invariants of
Hilsum--Skandalis Maps}, Ph.D Thesis, Univ.\ of Utrecht (1996).

\bibitem{Mrc2} J. Mr\v{c}un, \textit{
Functoriality of the bimodule associated to a Hilsum--Skandalis map},
 $K$-Theory \textbf{18} (1999), 
235--253.

\bibitem{MRW} P. Muhly, J. Renault, and D. Williams, \textit{Equivalence and 
isomorphism for groupoid \ca s}, J.\ Operator Th.\ \textbf{17} (1987), 3--22.


\bibitem{MS} P.S. Muhly and B. Solel, \textit{
 Tensor algebras over $C\sp *$-correspondences: representations,
 dilations, and $C\sp *$-envelopes},  J.\ Funct.\ Anal.\ \textbf{
158} (1998), 389--457.

\bibitem{Pas} W.L. Paschke,   \textit{Inner
product modules over $B^*$-algebras},  Trans.\ Amer.\ Math.\ Soc.\
{\bf 182} (1973),  443--468.

\bibitem{RW}
 I. Raeburn and D.P. Williams, \textit{Morita equivalence and
 continuous-trace $C\sp *$-algebras},
 American Mathematical Society, Providence, RI, 1998.

\bibitem{Ram1} A. Ramsay, \textit{Virtual groups and group actions},
Adv.\ Math.\ \textbf{6} (1971), 253--322.

\bibitem{Ram2} A. Ramsay, \textit{Topologies on measured groupoids},
J.\ Funct.\ Anal.\ \textbf{47} (1982), 314--343.

\bibitem{Ren} J. Renault,  \textit{A Groupoid Approach to $C^*$-algebras},
Lecture Notes in Mathematics \textbf{793}, Springer, Berlin, 1980.

\bibitem{Ren87} J.
Renault, \textit{Repr\'{e}sentation des produits crois\'{e}s
d'alg\`{e}bres de groupo\"{\i}des},  J.\ Operator Theory \textbf{18} (1987),
 67--97. 

\bibitem{Rie1} M.A. Rieffel, \textit{ Induced representations of
$C^*$-algebras},  Adv.\ Math.\ \textbf{13} (1974), 176--257.

\bibitem{Rie2} M.A. Rieffel, \textit{
Morita equivalence for $C^*$-algebras and
$W^*$-algebras},  J.\ Pure Appl.\ Alg.\ \textbf{5} (1974), 51--96.

\bibitem{Sau} J.-L.
Sauvageot, \textit{
Sur le produit tensoriel relatif d'espaces de Hilbert},
J.\ Operator Theory \textbf{9} (1983),  237--252. 

\bibitem{Sch} J. Schweizer, \textit{Crossed products by equivalence
bimodules}, Univ.\ T\"{u}bingen preprint (1999).

\bibitem{Vai} I. Vaisman,  \textit{Lectures on the Geometry of
Poisson Manifolds},  Birkh\"{a}user, Basel, 1994.


\bibitem{Was} A.
Wassermann, \textit{Operator algebras and conformal field
theory. III. Fusion of positive energy representations of ${\rm
LSU}(N)$ using bounded operators}, Invent.\ Math.\ \textbf{133} (1998),
467--538.

\bibitem{W0} A.
Weinstein, \textit{Lectures on symplectic manifolds}, Regional Conference
Series in Mathematics \textbf{29}, American Mathematical Society,
Providence, R.I., 1977.

\bibitem{W1} A.
Weinstein, \textit{The
 local structure of Poisson manifolds},   J.\ Diff.\ Geom.  \textbf{
 18} (1983), 523--557. Err.\ ibid.\ \textbf{22} (1985), 255.

\bibitem{W3} A. Weinstein, \textit{
Symplectic groupoids and Poisson manifolds}, 
Bull.\ Amer.\ Math.\ Soc.\ (N.S.) \textbf{16} (1987), 
101--104

\bibitem{W4} A. Weinstein, \textit{Coisotropic calculus and Poisson groupoids}, 
J.\ Math.\ Soc.\ Japan \textbf{40} (1988), 705--727.

\bibitem{W2} A.
Weinstein, \textit{Affine Poisson structures}, Int.\ J.\ Math.\ \textbf{1}
(1990), 343--360.

\bibitem{X1} P. Xu, \textit{ Morita equivalent symplectic
 groupoids}, pp.\ 291--311 in: P. Dazord and A.  Weinstein (eds.),
 \textit{Symplectic Geometry, Groupoids, and Integrable Systems},
 Springer, New York, 1991,
  
 \bibitem{X2} P. Xu, \textit{ Morita equivalence of Poisson
 manifolds}, Commun.\ Math.\ Phys.\ \textbf{142} (1991), 493--509.

\bibitem{X3}  P. Xu, 
   \textit{Morita equivalence and symplectic realizations of Poisson
manifolds}, Ann.\  Sci.\ \'{E}cole Norm.\ Sup.\  (4) \textbf{25} (1992),  307--333. 

\bibitem{Zak} S. Zakrzewski, \textit{Quantum and
 classical pseudogroups. I,II}, 
 Commun.\ Math.\ Phys. \textbf{134} (1990), 347--370,  371--395.

\end{thebibliography}
\end{document}